\newif\ifonecol 
\onecolfalse 

\ifonecol
\documentclass[journal,comsoc,12pt,onecolumn]{IEEEtran}
\else
\documentclass[journal,comsoc]{IEEEtran}
\fi

\newlength{\figurewidth}
\ifonecol
\else
\fi

\ifonecol
\usepackage{setspace}
\fi

\usepackage[T1]{fontenc}
\usepackage{color}

\ifCLASSINFOpdf
\else
\fi
\usepackage{amsmath}
\usepackage[cmintegrals]{newtxmath}

\usepackage{cancel}
\usepackage{soul}

\soulregister\cite7
\soulregister\ref7
\soulregister\pageref7
\usepackage[normalem]{ulem}

\usepackage{xcolor}

\newcommand{\fixme}[2]{\ifx&#2&{\color{red}#1}\else{\color{red}FIXME\{}#1{\color{red}\}}\footnote{{\color{red}#2}}\PackageWarning{Fixme}{#1: #2}\fi}

\usepackage[many]{tcolorbox}

\tcbset{
  highlight math style={
    colback=yellow,
    arc=0pt,
    outer arc=0pt,
    boxrule=0pt,
    top=2pt,
    bottom=2pt,
    left=2pt,
    right=2pt,
  }
}

\usepackage{mathtools}

\DeclarePairedDelimiter\floor{\lfloor}{\rfloor}

\newcommand{\MYfooter}{\smash{\scriptsize
\hfil\parbox[t][\height][t]{\textwidth}{\centering
\copyright 2020 IEEE. Personal use of this material is permitted. Permission from IEEE must be obtained for all other uses, including reprinting/republishing this material for advertising or promotional purposes, collecting new collected works for resale or redistribution to servers or lists, or reuse of any copyrighted component of this work in other works. DOI: 10.1109/TCOMM.2020.2987561.}\hfil\hbox{}}}

\makeatletter

\def\ps@IEEEtitlepagestyle{%
\def\@oddfoot{\MYfooter}%
\def\@evenfoot{\MYfooter}}

\makeatother

\begin{document}
%
\title{A Spatial Time-Frequency Hopping Index Modulated Scheme in Turbulence-free Optical Wireless Communication Channels}
%
%
%

\author{Francisco J. Escribano,~\IEEEmembership{Senior Member,~IEEE,} Alexandre Wagemakers, Georges Kaddoum,~\IEEEmembership{Member,~IEEE,} and Joao V. C. Evangelista,~\IEEEmembership{Student Member, IEEE}%
\thanks{Francisco J. Escribano is with the Signal Theory \& Communications Department, Universidad de Alcal\'{a}, 28805 Alcal\'{a} de Henares, Spain (email: francisco.escribano@ieee.org).}%
\thanks{Alexandre Wagemakers is with the Nonlinear Dynamics and Chaos Theory Group, Universidad Rey Juan Carlos, 28933 M\'{o}stoles, Spain (email: alexandre.wagemakers@urjc.es).}%
\thanks{Georges Kaddoum and Joao V. C. Evangelista are with the \'{E}cole de Technologie Sup\'{e}rieure, U. of Quebec, 1100 Notre-Dame St W, Montreal, Quebec H3C 1K3 Canada (e-mail: georges.kaddoum@etsmtl.ca).}}%

%
%

\markboth{Journal of \LaTeX\ Class Files,~Vol.~14, No.~8, August~2015}%
{Shell \MakeLowercase{\textit{et al.}}: Bare Demo of IEEEtran.cls for IEEE Communications Society Journals}
%



\maketitle

\begin{abstract}
In this article, we propose an index modulation system suitable for optical communications, where the intensity of the light is modulated along the time, space and frequency dimensions, building a Frequency-Hopping Spatial Multi-Pulse Position Modulation (FH-SMPPM). We analyze its performance from the point of view of its efficiency in power and spectrum, and its implementation complexity and required receiver latency. We analyze its error probability in the case of the non-turbulent free-space optical (FSO) channel. We derive formulas for the average symbol and bit error probabilities, and the simulation results show that they are tight enough for a wide range of signal-to-noise ratios and system parameters. We compare FH-SMPPM with other proposed index modulated systems of the same nature, and we highlight its distinctive advantages, like the flexibility to build an appropriate waveform under different constraints and objectives. Moreover, FH-SMPPM shows to be better performing in BER/SER and/or offer advantages in efficiency with respect to those alternatives, thus offering the possibility to be adopted for a variety of contexts.
\end{abstract}


%
\IEEEpeerreviewmaketitle

\section{Introduction}
%
%
%
%

\IEEEPARstart{T}{he} developments in 5G and beyond 5G witnessed nowadays impose very high throughput and extreme high quality requirements to be addressed at the PHY, in order to guarantee the successful deployment of the envisaged use cases and applications. Consequently, substantial efforts have been made to design innovative solutions which could potentially achieve the corresponding demands. In this context, the idea of index modulation (IM) systems \cite{8004416} is considered as one of the most promising proposals and has thus attracted significant attention. The concept of IM arises from the smart usage of the characteristics of the signals or systems entailed in a communication, tailored so as to carry additional information, which is codified in the specific setup, or through driving some specifically chosen waveform parameters.

Given their great potential, a variety of works have emerged on the subject. To date, diverse IM systems, in most cases directly associated with the infrastructure of the specific frontends, have been proposed and studied. For instance, spatial modulation (SM) was initially proposed in \cite{4382913}, where, in one of its alternatives, a Multiple-Input Multiple-Output (MIMO) system is configured to carry side information, codified in the chosen pattern of active antennas \cite{6166339}. Moreover, the idea of driving the physical elements of a set of reconfigurable antennas to further enhance the possibilities of IM for Radio Frequency (RF) transmission has also been studied \cite{6678765}. Other IM systems have also been designed under similar principles, by taking advantage of the several dimensions available in a MIMO framework, for example, through the so-called space-time-frequency shift keying \cite{5688440}. Under these strategies, the effects of dispersive channels can be counteracted by the diversity introduced in the system. For additional details, readers are directed to \cite{8417419}, which presents a tutorial on IM, discusses the state-of-the-art, and highlights some open challenges.

In the case of such MIMO-based solutions, the receiver usually requires prior knowledge as well as a constant update of the Channel State Information (CSI) over all the transmitting and receiving antenna paths. This imposes a substantial burden on the system, and such schemes have been shown to be very sensitive to noisy CSI estimations as well as to correlated channels \cite{chang2013detection,Soujeri2015performance}. Additionally, the switching time between antennas at the transmitter, i.e. the time needed for the RF switches to effectuate a transition among the transmitting antennas, is another limitation of some of these proposals, as it may easily result in a reduction of the system capacity \cite{soujeri2016impact}.

Another approach based on the idea of selecting between  different subcarrier patterns in Orthogonal Frequency-Division Multiplexing (OFDM) systems, denoted as OFDM-IM, has also led to the development of interesting alternatives \cite{6587554}. Such schemes are required to share an indexing Look-Up Table (LUT) between the communicating entities, and they could suffer from lack of flexibility. Furthermore, OFDM-IM receivers rely on maximum likelihood (ML) detection which requires performing an exhaustive search over all the possibilities of subcarrier combinations. Consequently, for large combination values, ML-based detectors become impractical due to the exponentially increasing decoding complexity needed \cite{6587554}. To overcome this issue, lower complexity detectors can be considered \cite{8477158}, including Log-Likelihood Ratio (LLR)-based detection strategies (as applied in the so-called dual-mode (DM)-OFDM \cite{7547943}), or the so-called Low-Complexity (LC)-OFDM-IM \cite{6587554, 7370760}. Very appealing is also the possibility to combine OFDM-IM and space-time shift keying (STSK) with the complexity reduction determined by compressed sensing \cite{8322306}, an idea further explored in \cite{8656571} for multi-dimensional IM. These systems trade-off final performance against reduced resources consumption.

Moreover, in order to achieve higher throughputs, another index modulation scheme, called Code Index Modulation-Spread Spectrum (CIM-SS), has been proposed in \cite{CIMTVT}. This system makes use of spreading codes jointly with the constellation symbols to map the data. At the receiver side, the modulated bits can be detected by first identifying the spreading code that yields the maximum autocorrelation value. Therefore, an increase in the number of mapped bits results in a shortage of spreading codes as well as an increasingly complex receiver implementation.  We can see that enhancing the communication capabilities through IM always leads to difficulties that have to be correspondingly addressed. Other recently investigated IM systems of this kind are based on chaos, like the combination of Pulse Position Modulation (PPM) and M-ary Differential Chaos Shift Keying (M-DCSK) \cite{8468068}, or the combination of M-DCSK and CIM \cite{8703732}. Their aim is to compensate for the poor distance properties of DCSK by mapping the symbol bits partly on the chaotic signal and partly on other properties of the waveform.

In any case, currently envisaged 5G developments are not restricted to RF, because Optical Wireless Communications (OWC) are also receiving increasing attention. The idea is that light can be a valid alternative to ground the PHY for small scale deployments, hence making use of its localized nature and non-penetrative characteristics to alleviate the current RF spectrum shortage, while easing the growing interference limitation concerns. Thus, similar to RF-based solutions, IM has been adopted in OWC systems, with a stress on both OFDM and MIMO applications \cite{7915761}, \cite{MAO201737}, \cite{8315127}, \cite{8765392}. For instance, Frequency Domain SM (FD-SM) and time domain SM (TD-SM) for OWC were investigated in \cite{8727949}. Moreover, Optical Spatial Modulation (OSM) is a MIMO technique that simultaneously conveys information through spatial and signal domains. In \cite{8732427}, a Differential-OSM (DOSM) system, which does not rely on prior knowledge of the channel state information at the receiver, was proposed.

In addition, IM systems specifically designed for OWC have led to the proposal of schemes that go beyond MIMO and OFDM, and attempt to make use of other physical features of the signals. For instance, in the context of single-carrier communications, the joint use of PPM or Frequency Shift Keying (FSK), while driving the phase or the polarization of the coherent light signal, was proposed in \cite{Liu:11}. The result is a compound symbol carrying information over different dimensions, that can thus increase the efficiency. Moreover, by considering multiple LEDs and photodiodes (PDs), the work in \cite{8642337} designed a low-complexity OSM scheme based on bipolar Pulse Amplitude Modulation (PAM) for MIMO Intensity Modulation/Direct Detection (IM/DD) OWC. In such scheme, the transmitting LEDs use space shift keying (SSK) to convey additional information. In yet other possibility, the use of Multi-PPM (MPPM) while simultaneously plugging additional information in the amplitude and phase dimensions of the active slots was proposed in \cite{6876375}, where a Quadrature Amplitude Modulation (QAM-)MPPM system is described. The disadvantage here is the requirement for coherent detection in the electrical domain.

In a similar perspective, Optical SSK (OSSK) and Spatial PPM (SPPM) have been proposed as variants of OSM. These schemes represent promising solutions for pulse-based OSM systems. For instance, OSSK is defined as a low-complexity OWC-adapted extension of RF Space Shift Keying (RF-SSK). OSSK works with incoherent light sources, and is thus based on IM/DD \cite{7858134}. In this scheme, the index of the pulsed LED conveys extra information, so that, for pure OSSK, $M$ LEDs can transmit $\log_2\left(M\right)$ bits per symbol duration. In the case of SPPM, additional bits per symbol can be transmitted by using both the active LED index and the pulse position index of the PPM constellation in order to map the bits \cite{6241395}. However, this comes at the cost of higher bandwidth occupancy and higher demodulator complexity.

On the other hand, in case of high loss channels and systems suffering from limited transmission power, i.e. in very low signal-to-noise ratio scenarios, FSK-based OWC systems present practical alternatives that yield promising error performances at the cost of some extra bandwidth expenditure \cite{Savage:13}. In this context, the modulation process is based on driving the wavelength of the emitted light according to FSK principles, and was consequently named Wavelength Shift Keying (WSK). A derivative of such systems, namely Color Shift Keying, was also proposed in \cite{6780585}. It is to be noted here that such systems require both light coherence and elaborated optical frontends.

Modulating the intensity of the light according to the waveform pattern of FSK is another way to make use of the advantages of frequency modulation in the optical domain. In the case where only a particular subcarrier is active at a given time (thus producing in practice a hopping pattern), such systems are considered to belong to the class of light intensity modulated optical OFDM systems, where we also find such alternatives as DC-biased Optical OFDM (DCO-OFDM) \cite{Gonzalez06} and Asymmetrically Clipped Optical OFDM (ACO-OFDM) \cite{1610439}. In these cases, light coherence is not a system requirement, and the demodulation performed in the electrical domain can be non-coherent. As a consequence, these possibilities yield simple and robust detectors. Hence, frequency-based systems of this kind are considered attractive for simple LED-based infrastructures, as they just require IM/DD.

Quite recently, other hybrid modulation schemes have emerged in the context of Visible Light Communications (VLC), where frequency- and time-domain, or frequency- and intensity-domain joint strategies are exploited, see \cite{8762056} and \cite{8819997}, respectively. Moreover, an IM system based on MPPM, where an FSK symbol is sent during the active slots, instead of just sending a constant amplitude rectangular pulse shape unable to carry any additional information, was proposed for OWC IM/DD communications in \cite{8720057}. This  system jointly drives the time and the frequency axis, and therefore constitutes an index-time frequency hopping (I-TFH) modulation.

Along these ideas, we propose in this work an IM system based on Spatial MPPM (SMPPM) \cite{8647138} that carries additional information during the signal slots by combining an FSK scheme, which can thus be named Frequency-Hopping Spatial Multi-Pulse Position Modulation (FH-SMPPM).  We will show how its spectral and power efficiencies, its demodulator complexity, and its error rate performance, can be efficiently traded-off respecting other related IM alternatives, like I-TFH \cite{8720057} and the mentioned SMPPM, which are limit cases of FH-SMPPM. Therefore, FH-SMPPM inherits comparative advantages and disadvantages of both systems, and can offer relative improvements with respect to pure I-TFH or pure SMPPM, depending on the specific scenario involved.

As will be detailed in the sequel, the main contributions of this paper can be summarized as
\begin{itemize}
 \item Proposal of the FH-SMPPM system and the corresponding demodulation scheme.
 \item Derivation of spectral and power efficiencies for FH-SMPPM.
 \item Derivation of estimated demodulator complexity and latency for FH-SMPPM.
 \item Derivation of tight average symbol and bit error probability analytical expressions for FH-SMPPM, that can also be applied to SMPPM.
 \item Illustrative comparisons among different possibilities and trade-offs in terms of spectral and power efficiency, of demodulator complexity, and of error rate performance.
\end{itemize}

The structure of the article is as follows. In Section \ref{model}, the FH-SMPPM signal model and its demodulation process are presented, along some of its characteristic parameters. Section \ref{analysis} presents the performance analysis of FH-SMPPM and its comparative alternatives respecting efficiency and complexity, and, correspondingly, the derivation of approximate expressions for the symbol and bit error probabilities. In Section \ref{results}, simulation results are provided which validate the tightness of the derived error probability approximations, and demonstrate the comparative advantages of FH-SMPPM. Finally, the conclusions are summarized in Section \ref{conclusions}.

\section{System model}
\label{model}

In this Section, we review and define the signals and meaningful parameters for the FH-SMPPM system. Along all this work, we will consider an i.i.d. binary source, and this means that each of the symbols involved in the different scenarios will be equiprobable. We consider an OWC setup with $M_S$ transmitters and one receiver, where $M_S$ is a power of $2$. The transmitters are located at different points in space, and they have adjustable transmitting powers, so that each optical communication link is set to have a different and distinctive gain, as in an OSSK system \cite{6241395}.

\begin{figure}[htb!]
\centering
\includegraphics[width=7.5cm]{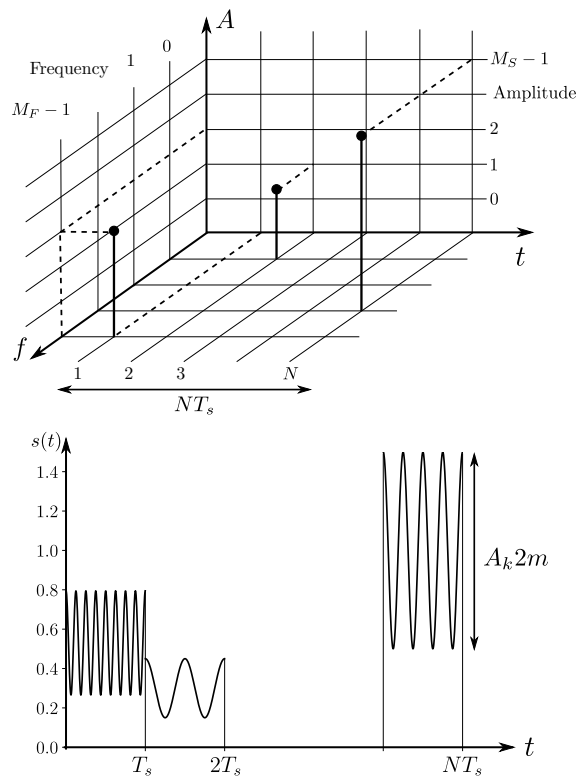}
 \caption{Signal model for FH-SMPPM in the time/frequency/amplitude frame, and its representation in the time domain.}
 \label{fig1}
\end{figure}
The transmitters are coordinated and they transmit following an MPPM signal pattern \cite {Hamkins2005MultipulsePM,Caplan2007}. The symbol has a period $T$, and it is divided into $N>1$ slots, with duration $T_s=T/N$ each. Within the interval $T$, only $w\in\left\{1,2,\cdots,N\right\}$ slots are used to send a nonzero signal from one (and just one per slot) of the transmitters. The nonzero slots are called signal slots, and the rest are called non-signal slots. The MPPM symbol part is therefore defined by an $N$-dimensional vector $\mathbf{d}$, belonging to the set
\begin{equation}
 \mathcal{S}_{\mathrm{MPPM}}=\left\{ \mathbf{d}=\left(d_0,\cdots,d_{N-1}\right) \in \{0,1\}^N : \sum_{k=0}^{N-1} d_k=w\right\}.
\end{equation}
Accordingly, the number of bits per MPPM symbol will be
\begin{equation}
q_{\mathrm{MPPM}}=\floor*{\log_2 {\binom{N}{w}}},
\end{equation}
which is maximum for $w=\floor*{N/2}$. Notice that we only use $2^{q_{\mathrm{MPPM}}} \leq \binom{N}{w}$ MPPM symbols from the set $\mathcal{S}_{\mathrm{MPPM}}$: the expurgated set with the actual MPPM symbols will be denoted as $\mathcal{S}_{\mathrm{MPPM}}^*$.

The information codified in the OSSK symbol corresponds to $n_S=\log_2\left(M_S\right)$ bits, because each of the transmitters is identified by a unique bit pattern. As we have a total of $w$ signal slots, the number of bits contained in the OSSK symbol part is
\begin{equation}
q_{\mathrm{OSSK}}=w \log_2\left(M_S\right)=w n_S.
\end{equation}
The possibility to communicate using this kind of compound symbol has already been studied and evaluated in a system denominated SMPPM \cite{8647138}.

As a way to extend the possiblities of the SMPPM symbol, and make it carry additional information, instead of just sending rectangular optical pulses within each of the $w$ signal slots, another kind of modulated symbol may be sent by driving other properties of the signal. Similar ideas have already been exploited, for example by using PPM and jointly driving the phase or polarization of the optical signal \cite{Liu:11}, or by using MPPM and jointly driving the frequency of the light intensity fluctuations during the nonzero transmitted optical pulses \cite{8720057}.

We resort to the last mentioned strategy, and additional information bits are mapped during the signal slots by using an FSK scheme, with $M_F$ (a power of $2$) available frequencies. The number of FSK bits per FH-SMPPM symbol will thus be
\begin{equation}
 q_{\mathrm{FSK}}=w \log_2\left(M_F\right)=w n_F,
\end{equation}
where we have defined $n_F=\log_2\left(M_F\right)$. The total number of bits per FH-SMPPM symbol is
\begin{equation}
 q_{\mathrm{F-S}}=w \left(n_S + n_F\right) + \floor*{\log_2 {\binom{N}{w}}},
\end{equation}
and the binary rate is $R_b=q_{\mathrm{F-S}}/T=q_{\mathrm{F-S}}/\left(NT_s\right)$.

With these ideas, we can now describe the waveform in the time domain. If we want to consider the optical communications (OC) channel, we have to make sure that the waveform takes only positive values, given that in the simplest OC general case (e.g. non-coherent LED- or laser-based communications of any kind), the transmission would be made using intensity modulation, and the reception at the photodiode (PD) would be made through direct detection (standard low-complexity IM/DD schemes). In this case, we may write the electrical waveform, in a symbol period $0\leq t < T$, as
\begin{equation}
\label{wave}
s\left( t \right) = \sum_{k=0}^{N-1} A_k d_k p \left(\frac{t-k T_s}{T_s} \right) \biggl[ 1+ m \cos\left( 2 \pi f_k t\right) \biggr],
\end{equation}
where $d_k$ are the vector components of $\mathbf{d}$ for the MPPM symbol part, $p\left( t \right)$ is the unit-duration unit-amplitude rectangular pulse,
\begin{equation}
 \label{rec_pulse}
 p\left(t\right) = \left\{\begin{array}{ll} 1, & 0\leq t \leq 1 \\ 0, & \text{otherwise} \end{array}\right. ,
\end{equation}
$0< m \leq 1$ is a modulation index, $f_k$ is the FSK symbol frequency, so that
\begin{equation}
 f_k= \left\{ \begin{array}{ll} 0, & d_k=0 \\ n_{i\left(k\right)}/T_s, & d_k \neq 0  \end{array}\right. ,
\end{equation}
where $n_{i\left(k\right)} \gg 1$ is a positive integer defining the specific FSK frequency in the corresponding signal slot ($i\left(k\right) = 0,1,\cdots,M_F-1$ denotes thus the index of the frequency for the $k$-th signal slot),  and $A_k$ is the amplitude of the pulse, so that
\begin{equation}
 A_k= \left\{ \begin{array}{ll} 0, & d_k=0 \\ T_{j\left(k\right)}, & d_k \neq 0  \end{array}\right. ,
\end{equation}
where $T_{j\left(k\right)}$ is the transmitting amplitude of the $j\left(k\right)$-th activated OSSK transmitter during the $k$-th signal slot (and therefore $j\left(k\right) = 0,1,\cdots,M_S-1 $). We choose the frequencies of the FSK setup so that the separation among adjacent ones is
\begin{equation}
\Delta f=\frac{1}{T_s},
\end{equation}
for minimum bandwidth usage, as we will resort to non-coherent demodulation of the FSK symbols recovered in the electrical domain. We denote as $\mathcal{S}_{\mathrm{FSK}}$ the set of $M_F$ frequencies in the FSK scheme.

The signal defined in \eqref{wave} contains a bias in the signal slots to guarantee that it does not experience clipping at the optical interface. In Fig. \ref{fig1} we can see a depiction of an actual FH-SMPPM symbol in the time domain, and of its representation in the time/frequency grid. As is done in already well-known proposals resorting to IM/DD \cite{6876375}, \cite{Gonzalez06}, \cite{1610439}, we assume this electrical signal is linearly converted into a light intensity waveform.

\begin{figure*}
\centering
\includegraphics[width=0.9\textwidth]{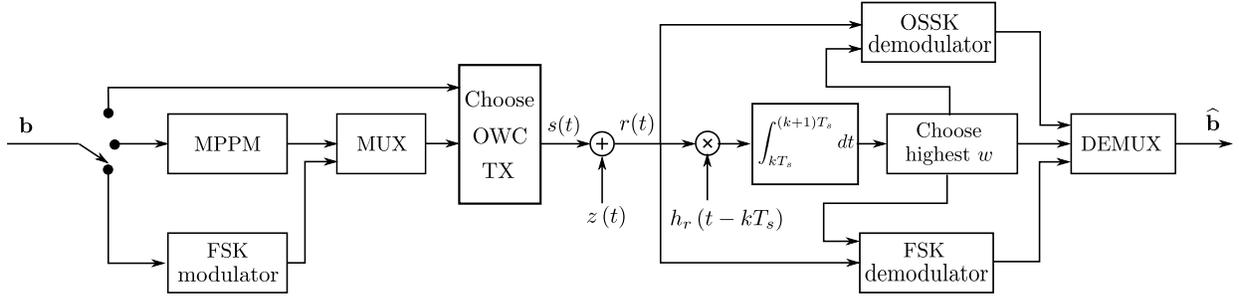}
 \caption{System model for FH-SMPPM transmitter, channel and receiver. The TX MUX block plugs the FSK symbols into the signal slots of the MPPM frame, while the RX DEMUX block maps the separately demodulated bits into the recovered information bit stream. The matched impulse response $h_r\left(t\right)$ is given in \eqref{norm_rec_pulse}.}
 \label{fig2}
\end{figure*}
\subsection{Detection of FH-SMPPM}
As shown in Fig. \ref{fig2}, the received signal in the electrical domain after the transduction at the PD\footnote{Under the typical hypothesis in the field of IM/DD communications, its output signal would contain a term proportional to the light intensity waveform impinging the detector.} can be modeled as
\ifonecol
\begin{equation}
\label{rx}
 r\left(t\right)=\sum\limits_{k=0}^{N-1} R_k d_k p\left(\frac{t-kT_s}{T_s} \right) \biggl[ 1 + m \cos \left( 2 \pi f_k t + \theta_k \right)\biggr] +z\left( t \right),
\end{equation}
\else
\begin{eqnarray}
\label{rx}
 &  r\left(t\right)=\sum\limits_{k=0}^{N-1} R_k d_k p\left(\frac{t-kT_s}{T_s} \right) \biggl[ 1 + m \biggr.  \nonumber \\
 & \biggl. \cdot \cos \left( 2 \pi f_k t + \theta_k \right)\biggr] +z\left( t \right),
\end{eqnarray}
\fi
where $\theta_k$ is an unknown random phase that accounts for the non-coherent reception\footnote{Note that, given the model described, we have to understand coherence in this context exclusively in the electrical domain.}, $z\left(t\right)$ is an instance of real-valued additive white Gaussian noise with power spectral density $N_0/2$, and $R_k$ is the PD average output electrical current during the $k$-th signal slot, so that
\begin{equation}
 R_k= \left\{ \begin{array}{ll} 0, & d_k=0 \\ H_{j\left(k\right)}, & d_k \neq 0  \end{array}\right. ,
\end{equation}
where, in the case the $j\left(k\right)$-th transmitter is active,
\begin{equation}
 H_{j\left(k\right)}=T_{j\left(k\right)} \mathcal{R} G_{j\left(k\right)},
\end{equation}
where $G_{j\left(k\right)}$ is the optical channel gain for the path between $j\left(k\right)$-th transmitter and the receiver at the $k$-th signal slot, and $\mathcal{R}$ is the responsivity of the PD. The optical channel gain for each path is considered constant in the case of non-turbulent FSO channels, or time-variant in the case of turbulent FSO channels. In this article, we consider a constant non-turbulent FSO channel, and, as the output amplitude and channel gain would be constant for each OSSK transmitter, in the sequel we drop the dependence with $k$ in our discussions about the OSSK part.

Notice that, for the OSSK symbol part to work properly, we have to make sure to tune amplitudes $T_j$ so that $H_j \neq H_l$ for $j \neq l$ \cite{6241395}. An even distribution of amplitudes will be the most advantageous case, since this maximizes the minimum distance in the OSSK symbol part, and we therefore assume that the amplitudes are chosen so that
\begin{equation}
 H_j = I_m \left( 1 - L_m \frac{j}{M_S-1}\right), \, j=0,\cdots,M_S-1,
\end{equation}
where $I_m$ is the maximum possible average current in a signal slot, and $0 < L_m <1$ is a limiting factor so that the minimum average current is larger than $0$. We denote as $\mathcal{S}_{\mathrm{OSSK}}$ the set of $M_S$ received amplitudes in the OSSK setup. The mapping from bits to symbols follows the Gray coding scheme, according to the value of the received amplitudes.

The average received optical power is proportional to the DC value of the received signal current, and it is given by
\begin{equation}
 \label{arop}
 P_{opt}=\frac{I_{DC}}{\mathcal{R}}=\frac{w}{N} \frac{I_{ph}}{\mathcal{R}},
\end{equation}
where the average photocurrent within a signal slot $I_{ph}$ is given by
\begin{equation}
 I_{ph} = \mathrm{E}\left[ H_j \right] = \frac{1}{M_S}\sum_{j=0}^{M_S-1} H_j= I_m \left(1 - \frac{L_m}{2} \right).
\end{equation}
The electrical average received symbol energy is
\ifonecol
\begin{equation}
\label{Es}
 E_{s,\mathrm{F-S}}= w T_s \mathrm{E} \left[ H_j^2 \right]  \left(1+\frac{m^2}{2}\right) = w T_s I_m^2 \left(1 + L_m^2 \left(\frac{1}{3}+\frac{1}{6\left(M_S-1\right)}\right) + L_m \right) \left(1+\frac{m^2}{2}\right),
\end{equation}
\else
\begin{eqnarray}
\label{Es}
 & \displaystyle E_{s,\mathrm{F-S}}= w T_s \mathrm{E} \left[ H_j^2 \right]  \left(1+\frac{m^2}{2}\right) = w T_s I_m^2 &\\
 & \displaystyle \cdot \left(1 + L_m^2 \left(\frac{1}{3}+\frac{1}{6\left(M_S-1\right)}\right) - L_m \right) \left(1+\frac{m^2}{2}\right), \nonumber
\end{eqnarray}
\fi
and the squared minimum distance between received FH-SMPPM symbols is
\begin{equation}
\label{dmin}
 d_{\mathrm{min, F-S}}^2=\min \left(T_s I_m^2 \left(\frac{L_m}{M_S-1}\right)^2 , T_s  I_m^2 \left(1-L_m\right)^2 m^2\right),
\end{equation}
that clearly depends on the chosen values for $m$, $L_m$ and $M_S$. The first possibility within the $\min\left(\cdot,\cdot\right)$ function  corresponds to the case of considering a pair of symbols where there is just a difference in the amplitude values for a single signal slot, and the second possibility corresponds to the case where the only difference resides in the frequencies of the FSK part within a single signal slot. When $L_m$ is near to $1$, the minimum distance will be given normally by the FSK symbol part (second possibility), and, when $L_m$ is near to $0$, the minimum distance will be determined by the OSSK symbol part (first possibility).

For the noise $z\left(t\right)$, we choose a standard model \cite{RoFT02,AOL04}, where its unilateral power spectral density at the optical receiver can be calculated as
\begin{equation}
\label{N0}
 N_0=\frac{4 k_B T_R F}{R_L} + 2 \left| q \right| I_{DC} + (RIN) I_{DC}^2 ,
\end{equation}
where $k_B$ is the Bolztmann constant, $T_R$ is the reference absolute temperature, $F$ is the receiver electronics noise factor, $R_L$ is the PD load resistor, $q$ is the electron charge, and $(RIN)$ is the relative-intensity noise factor. The first term on the RHS is the thermal noise, the second the shot noise, and the third, the relative-intensity noise (RIN).

The demodulation of FH-SMPPM will be made in two steps. In a first step, for each time slot, $r\left(t\right)$ is correlated with the normalized rectangular pulse shape matched impulse response
\begin{equation}
 \label{norm_rec_pulse}
 h_r\left(t\right) = \left\{\begin{array}{ll} \frac{1}{\sqrt{T_s}}, & 0\leq t \leq T_s \\ 0, & \text{otherwise} \end{array}\right. ,
\end{equation}
so as to get the metric
\begin{equation}
 \label{metric_MPPM}
 r_k = \int_{kT_s}^{\left(k+1\right)T_s} h_r\left(t-kT_s\right) r\left(t\right) dt = \sqrt{T_s} R_k d_k + n_k,
\end{equation}
where $n_k$ is a zero-mean Gaussian random variable (RV) with variance $N_0/2$. As detailed in \cite{8647138}, the hypothetical signal slots are identified as the $w$ slots with highest $r_k$ values (this allows the recovery of the corresponding $q_{\mathrm{MPPM}}$ bits mapped in the MPPM symbol part). After this, the ML OSSK detector is applied to the selected $w$ signal slots over the metrics $r_k$, so as to get the corresponding $q_{\mathrm{OSSK}}$ bits. Notice that the same metric is used to detect both the MPPM and the OSSK symbol parts. We can also see that the random variable (RV) $X_k=r_k$ defined by this metric is a zero-mean Gaussian RV with variance $N_0/2$ in the case of the non-signal slots, and is a Gaussian RV with mean $\sqrt{T_s} H_j$ and variance $N_0/2$ in the case of the signal slots, when the $j$-th transmitter was active.

To demodulate the FSK symbol part, we consider a detection process like the one described in \cite{8720057}, where the standard non-coherent FSK receiver is applied over the $w$ MPPM signal slots identified during the previous phase. The FSK detector calculates, for the hypothetical $k-$th signal slot and $i=0,\cdots,M_F-1$,
\begin{eqnarray}
 &r_{ik}^{\mathrm{I}}=\displaystyle\int_{kT_s}^{\left(k+1\right)T_s} r\left( t \right) \sqrt{\frac{2}{T_s}} \cos\left(2 \pi f_i t \right) dt, \nonumber \\
 &r_{ik}^{\mathrm{Q}}=\displaystyle\int_{kT_s}^{\left(k+1\right)T_s} r\left( t \right) \sqrt{\frac{2}{T_s}} \sin\left(2 \pi f_i t \right) dt.
\end{eqnarray}
The set of metrics
\begin{equation}
\label{metric_FSK}
 X_{ik}=\bigl| r_{ik}^{\mathrm{I}}\bigr|^2 + \bigl| r_{ik}^{\mathrm{Q}} \bigr|^2
\end{equation}
are used to demodulate the FSK symbol: the highest value over $i$ will determine the hypothesis about the frequency that has more likely been sent (and we thus get the $q_{\mathrm{FSK}}$ mapped bits). Notice that, as a difference from \cite{8720057}, the energy of the FSK symbol depends on the transmitter active during the corresponding signal slot. If the $j$-th transmitter was active, the energy of the FSK symbol part would be
\begin{equation}
\label{E_FSK}
 E_{s,\mathrm{FSK}}\left(H_j\right) = T_s H_j^2 \frac{m^2}{2}.
\end{equation}
Notice that the demodulation of FSK, as with OSSK, is not independent from the demodulation of MPPM.

\section{System analysis}
\label{analysis}

In this section we will address the analysis of the FH-SMPPM system, from the point of view of the efficiency and complexity of the scheme, and from the point of view of the final error performance.

\subsection{Efficiency analysis}
\label{efficiency}

\begin{figure}
\centering
\includegraphics[width=\figurewidth, keepaspectratio]{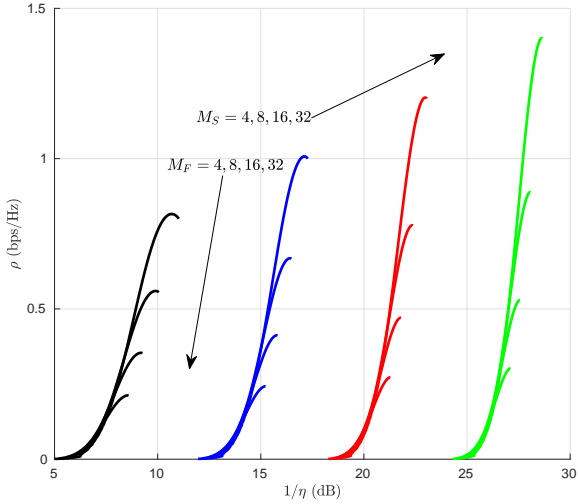}
\caption{Spectral efficiency ($\rho$) against the inverse of the asymptotic power efficiency in dB ($-10\log_{10}\left(\eta\right)$), for different cases of interest. Data have been generated with $m=0.9$, $I_m=1$, $L_m=0.7$, $N=1,\cdots,512$ and $w=1,\cdots,N$. $M_S$ and $M_F$ sweep values from $4$ to $32$.}
\label{fig3}
\end{figure}
We can compare the communication systems from the point of view of the spectral efficiency
\begin{equation}
 \rho \triangleq \frac{R_b}{B},
\end{equation}
where $R_b$ is the binary rate of the system and $B$ the occupied bandwidth. Given that we are using rectangular pulse shaping, and a slot period of $T_s$, the occupied bandwidth will be $\left(M_F+1\right) \Delta f =\left( M_F+1 \right) / T_s$ when we resort to minimal frequency separation for non coherent FSK. Therefore
\begin{equation}
 \rho_{\mathrm{F-S}}=\frac{q_{\mathrm{F-S}}}{N\left(M_F+1\right)}=\frac{w \left(n_S+n_F\right)+\floor*{\log_2 {\binom{N}{w}}}}{N\left(M_F+1\right)}.
\end{equation}

\begin{figure}
\centering
\includegraphics[width=\figurewidth, keepaspectratio]{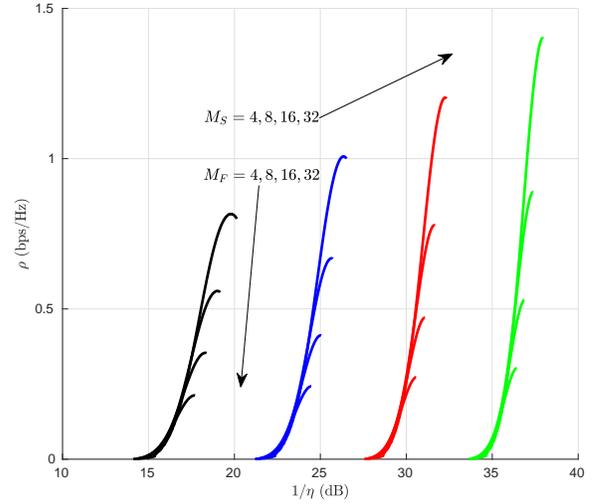}
\caption{Spectral efficiency ($\rho$) against the inverse of the asymptotic power efficiency in dB ($-10\log_{10}\left(\eta\right)$), for different cases of interest. Data have been generated with $m=0.9$, $I_m=1$, $L_m=0.3$, $N=1,\cdots,512$ and $w=1,\cdots,N$. $M_S$ and $M_F$ sweep values from $4$ to $32$.}
\label{fig4}
\end{figure}
The asymptotic power efficiency  \cite{5191053}, \cite{7067350} is defined as
\begin{equation}
 \eta \triangleq \frac{d_{\mathrm{min}}^2 \log_2\left(M_{\mathrm{sym}}\right)}{4 E_s},
\end{equation}
where $d_{\mathrm{min}}^2$ is the minimum square distance of the constellation, $M_{\mathrm{sym}}$ is the cardinality of the symbol set, and $E_s$ is the average symbol energy. Therefore
\begin{equation}
 \displaystyle \eta_{\mathrm{F-S}}=\frac{\min\left(\left(\frac{L_m}{M_S-1}\right)^2, \left(1-L_m\right)^2m^2\right) q_{\mathrm{F-S}}}{4w\left(1+L_m^2\left(\frac{1}{3}+\frac{1}{6\left(M_S-1\right)}\right)-L_m\right)\left(1+\frac{m^2}{2}\right)}.
\end{equation}

In Figs. \ref{fig3} and \ref{fig4} we can see the plane $\rho$-$1/\eta$ for different cases of interest. When the span $L_m$ is higher ($L_m=0.7$ in Fig. \ref{fig3}, and $L_m=0.3$ in Fig. \ref{fig4}), the assymptotic power efficiency is higher for a given set of parameters, and the curves are shifted to the left. On the other hand, we see the common trends with frequency-based modulations against amplitude-based modulations: when $M_S$ grows, the spectral efficiency grows, but the power efficiency decreases; when $M_F$ grows, the spectral efficiency decays, but the power efficiency increases. Changing the parameters of the MPPM symbol part also shifts the working point over each of the curves depicted for fixed $M_S$ and $M_F$. These facts point out the trade-offs to be taken into account when setting the defining parameters of the FH-SMPPM waveform, whose choice will depend on the specific scenario.

\subsection{Computational complexity and processing  latency}
\label{comp}

To evaluate the computational complexity of the FH-SMPPM receiver structure, we have to take into account that the demodulating operations will take place in the discrete-time sampled domain. The minimum sampling frequency $f_s$ required to demodulate the signal depends directly on the bandwidth required by the transmission of the non-coherent FSK symbols, which is about $(M_F+1)/T_s$. If we apply the Nyquist-Shanon sampling theorem we get
\begin{equation}
f_s > 2 \frac{M_F+1}{T_s}.
\end{equation}
This inequality sets a lower bound for the number of samples per symbol $T_s f_s > 2\cdot (M_F+1)$. We choose $N_s=T_s f_s =  4\cdot (M_F+1)$ for the purposes of the present study. Therefore, the main contributions to the complexity when demodulating the compound IM symbol can be summarized as:
\begin{itemize}
 \item The matched filter to detect the MPPM pulses has a complexity of $\mathcal{O}\left( N N_s\right)=\mathcal{O}\left(4 N  (M_F+1)\right)$.
 \item The sorting algorithm to demodulate the MPPM symbol has at most a complexity of $\mathcal{O}\left(N\log w\right) + \mathcal{O}\left(1\right)$.
 \item The FSK filters for non-coherent detection have a complexity of $\mathcal{O}\left(2 M_F N_s w\right)=\mathcal{O}\left(8 M_F(M_F+1) w\right)$.
 \item The sorting algorithm to demodulate the FSK symbol has a complexity of $\mathcal{O}\left(M_F w\right)$.
 \item The sorting algorithm to demodulate the OSSK symbols has a complexity of $\mathcal{O}\left(M_S w\right)$.
\end{itemize}
There are thus four parameters that will determine the overall complexity, namely $w$, $N$, $M_F$, and $M_S$.

The MPPM detection can be decomposed in two parts, as illustrated in Fig. \ref{fig2}. The first step is to detect the $w$ signal slots with the highest energy after the matched filter. This search task can be done with a heap structure with  complexity $\mathcal{O}\left(N\log w\right)$. The second task is to match the MPPM codeword to its corresponding bit sequence. This can be done in constant time by using LUTs that include all possible $2^N$ MPPM patterns (not only the chosen ones). For high values of $N$ the LUT size explodes and more efficient techniques are needed \cite{siyu2009,liu2012}. The decoding complexity of FSK depends on the square of the number of frequencies $M_F$, while the OSSK complexity is linear with the number of transmitters $M_S$. Since these two parameters are powers of 2, the complexity of the corresponding decoders increases exponentially with the respective number of coded bits.
\begin{table}
\begin{center}
  \begin{tabular}{|c|c|c|c|}
  \hline
    & $M_F=4$  & $M_F=8$ &   $M_F=16$  \\
    & $M_S=16$ & $M_S=8$  & $M_S=4$ \\
  \hline
  FSK  &  1312 & 4672 & 17536 \\
  \hline
  OSSK   & 128& 64 & 32 \\
  \hline
  Matched filter &  320 & 576 & 1088\\
  \hline
  MPPM  & \multicolumn{3}{|c|}{ 35 }  \\
  \hline
  \end{tabular}
   \vspace*{0.2cm}
\caption{Estimation of the number of decoding operations for different FSK frequencies ($M_F$) and different number of transmitters ($M_S$). The parameters of the MPPM scheme are $N=16$ and $w=8$.}
\label{Tab1}
\end{center}
\end{table}
The FSK decoding will take most of the demodulation computing load, as can be seen in Table \ref{Tab1} for different combinations of $M_F$ and $M_S$.

Another meaningful figure of merit to be taken into account for a possible implementation is the  latency, measured between the arrival time of the signal samples at the demodulator input, and the availability of the detected data at the demodulator output. We can provide both a lower bound and a reasonable upper bound for this latency, if we assume some simple and quite general design hypothesis. The theoretical lower bound for the latency is $t_l = N\cdot T_s$ which is the duration of the FH-SMPPM symbol. However, the actual processing of the complete FH-SMPPM frame will take some more time, so that this is an unrealistic limit. Since the structure and complexity of the decoder does not depend on the slot period $T_s$, it is convenient to normalize the latency into a non-dimensional quantity, namely $\tau_l = N$.

Considering an implementation in modern hardware platforms, such as FPGA or DSP, the samples corresponding to each of the slots can be processed independently through the filtering units at the same time the FH-SMPPM symbol is actually being received. After buffering the samples of the $k$-th slot, the metrics in equation \eqref{metric_MPPM} and equation \eqref{metric_FSK} can be calculated during the buffering of the $k+1$-th slot. In doing so, all the metrics would be available shortly after the last sample of the symbol has been received. All the other operations have to be executed after the filtering operation of the $j$-th FH-SMPPM symbol, and should be finished before the $j+1$-th FH-SMPPM symbol has been completely received. For a processor that cannot execute the filtering independently from the other processing operations, such as the MPPM decoding, we have an upper bound for the latency of $\tau_l = 2 N$. If the architecture is able to operate in parallel mode, the whole processing has to be completed during the first slot of the $j+1$-th FH-SMPPM symbol. This sets an upper bound of $\tau_l = (N+1)$, which results in an important improvement, although it imposes stricter constrains in the hardware implementation.

To provide an idea of the relative importance of each parameter in the computational effort, we may assume a naive implementation of each processing block, characterized by a number of operations equal to the worst case (as detailed in the previous complexity analysis). We should however take first a decision on whether to provide an implementation focused on lower latency or on lower complexity. With the choice of the lower latency $\tau_l = (N+1)$ implementation, the complexity of the FSK demodulator is slightly different as it must compute the metrics for all the $N$ slots, and not exclussively for the $w$ most probable signal slots. The modified complexity is then $\mathcal{O}\left(8 M_F(M_F+1) N\right)$.
If we normalize the total number of operations with respect to the number of bits per FH-SMPPM symbol, we can write the number of operations per decoded bit as
\ifonecol
\begin{equation}
\displaystyle \eta_{\textrm{ops-per-bit}} = \frac{4 N  \left(M_F+1\right) + N\log w  + 8 M_F \left(M_F+1\right) N+ M_F w + M_S w }{w \left(n_S + n_F\right) + \floor*{\log_2 {\binom{N}{w}}}}.
\end{equation}
\else
\begin{eqnarray}
&\displaystyle \eta_{\textrm{ops-per-bit}} = & \\
&\displaystyle \frac{\!4N\! \left(M_F\!+\!1\right)+ N\log w+ 8 M_F\! \left(M_F\!+\!1\right)N + M_F w+M_S w }{w \left(n_S + n_F\right) + \floor*{\log_2 {\binom{N}{w}}}}.& \nonumber
\end{eqnarray}
\fi
\begin{figure}[htb!]
\begin{center}
\includegraphics[width=\figurewidth, keepaspectratio]{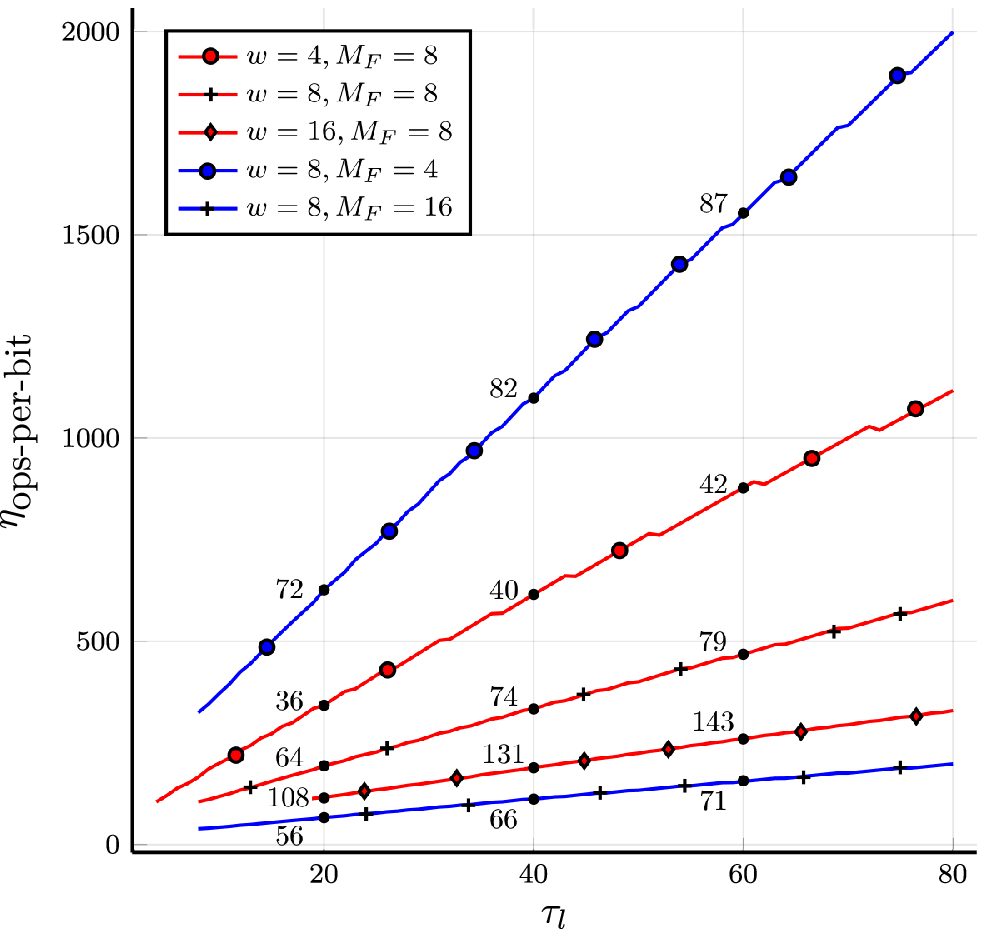}\\(a)\\
\includegraphics[width=\figurewidth, keepaspectratio]{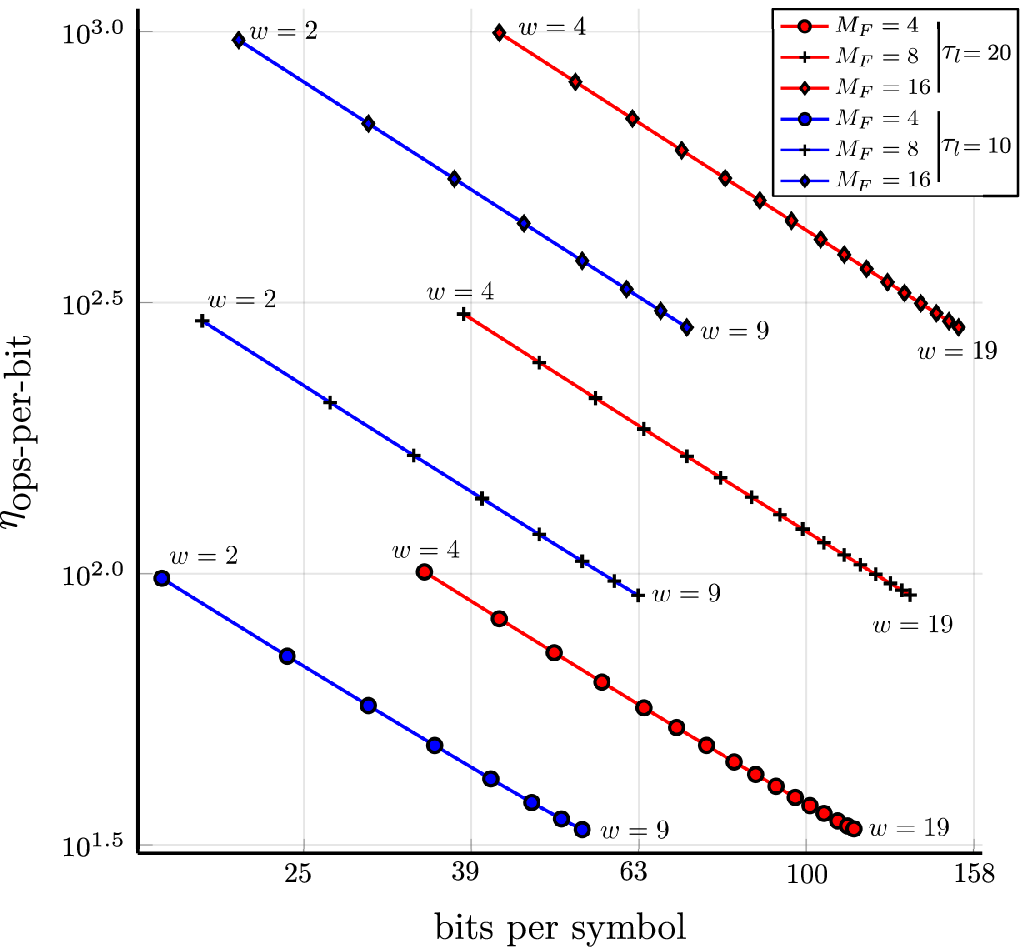}\\(b)
\end{center}
\caption{(a) Number of operations per decoded bit against the decoding latency $\tau_l = N+1$. The labels on the curves indicate the number of bits per symbol for some characteristic points. (b) Number of operations per bit as a function of the number of bits per symbol in a log-log plot. $M_S=8$ for all the curves.}
\label{fig5}
\end{figure}

In Fig. \ref{fig5} (a) we plot $\eta_{\textrm{ops-per-bit}}$ against the estimated latency for different parameters: we can thus visualize the possible trade-offs among all the parameters. These curves help to illustrate the balance between the latency and the resources consumption at the demodulator. For the chosen configuration, the number of operations per bit grows almost linearly with the latency as shown in Fig. \ref{fig5} (a). The number of bits per symbol is depicted at some specific points on the graph, in order to get a clear picture of the situation. If the latency is the most important design restriction, these curves show that a convenient choice of the number of occupied slots $w$ and the number of FSK frequencies $M_F$ can influence the bit rate and the complexity. However these choices may also be constrained by the spectral or power efficiencies. Figure \ref{fig5} (b) presents in a log-log plot the computational cost per bit $\eta_{\textrm{ops-per-bit}}$ as a function of the number of bits per symbol. In this case, if the bit rate is the important parameter, the choice of the other parameters ($\tau_l$, $w$ and $M_F$) allows to tune the complexity according to the design goals. Interestingly, these curves show that for larger delays not all possible bit rates are allowed.

Notice that in this study we have assumed an unlimited amount of memory available for the storage of the received samples, and for the MPPM LUT. Since the memory availability will strongly depend on the specific hardware chosen, we leave further considerations to the interested reader.

\subsection{Error performance analysis}
\label{error}

\subsubsection{Average symbol error probability}
\label{ASEP}

The average symbol error probability can be readily calculated by averaging over the conditional symbol error probability. An FH-SMPPM symbol is defined by the specific MPPM pattern $\mathbf{d} \in \mathcal{S}^*_{\mathrm{MPPM}}$, the specific FSK frequency pattern $\left\{ f_{i_l} \right\}_{\mathbf{I}} \in \left(\mathcal{S}_{\mathrm{FSK}}\right)^w$, where $\mathbf{I}=\left\{i_0,\cdots,i_{w-1}\right\} \in \mathcal{P}_w^{M_F}$ and  $\mathcal{P}_w^{M_F}$ is the set of all permutations with repetition of $w$ indexes taking values from $0$ to $M_F-1$, and the specific OSSK amplitude pattern $\left\{ H_{j_n} \right\}_{\mathbf{J}} \in \left(\mathcal{S}_{\mathrm{OSSK}}\right)^w$, where $\mathbf{J}=\left\{j_0,\cdots,j_{w-1}\right\} \in \mathcal{P}_w^{M_S}$ and  $\mathcal{P}_w^{M_S}$ is the set of all permutations with repetition of $w$ indexes taking values from $0$ to $M_S-1$. In these conditions
\ifonecol
\begin{equation}
 \label{Pe_avg}
 \displaystyle P_e = \mathrm{E}\left[P_e\left(\left\{ H_{j_n} \right\}_{\mathbf{J}},\left\{ f_{i_l} \right\}_{\mathbf{I}},\mathbf{d}\right)\right]= \frac{1}{2^{q_{F-S}}} \sum_{\mathbf{J} \in \mathcal{P}_{w}^{M_S}} \sum_{\mathbf{I} \in \mathcal{P}_{w}^{M_F}} \sum_{\mathbf{d} \in \mathcal{S}^*_{\mathrm{MPPM}}} P_e\left(\left\{ H_{j_n} \right\}_{\mathbf{J}},\left\{ f_{i_l} \right\}_{\mathbf{I}},\mathbf{d}\right),
\end{equation}
\else
\begin{eqnarray}
 \label{Pe_avg}
 & \displaystyle P_e =  \mathrm{E}\left[P_e\left(\left\{ H_{j_n} \right\}_{\mathbf{J}},\left\{ f_{i_l} \right\}_{\mathbf{I}},\mathbf{d}\right)\right]& \\
 & \displaystyle = \frac{1}{2^{q_{F-S}}} \sum_{\mathbf{J} \in \mathcal{P}_{w}^{M_S}} \sum_{\mathbf{I} \in \mathcal{P}_{w}^{M_F}} \sum_{\mathbf{d} \in \mathcal{S}^*_{\mathrm{MPPM}}} P_e\left(\left\{ H_{j_n} \right\}_{\mathbf{J}},\left\{ f_{i_l} \right\}_{\mathbf{I}},\mathbf{d}\right),& \nonumber
\end{eqnarray}
\fi
where $P_e\left(\left\{ H_{j_n} \right\}_{\mathbf{J}},\left\{ f_{i_l} \right\}_{\mathbf{I}},\mathbf{d}\right)$ is the conditional symbol error probability under the hypothesis of having sent a specific FH-SMPPM symbol. This conditional symbol error probability can be calculated as one minus the probability of correct detection, $P_c\left(\left\{ H_{j_n} \right\}_{\mathbf{J}},\left\{ f_{i_l} \right\}_{\mathbf{I}},\mathbf{d}\right)$, which can be factorized as
\ifonecol
\begin{eqnarray}
 \label{Pc_cond}
 & P_c\left(\left\{ H_{j_n} \right\}_{\mathbf{J}},\left\{ f_{i_l} \right\}_{\mathbf{I}},\mathbf{d}\right) = P_{c,\mathrm{MPPM}}\left(\left\{ H_{j_n} \right\}_{\mathbf{J}},\left\{ f_{i_l} \right\}_{\mathbf{I}},\mathbf{d}\right) & \\
 & \cdot P_{c,\mathrm{OSSK}}\left(\left\{ H_{j_n} \right\}_{\mathbf{J}},\left\{ f_{i_l} \right\}_{\mathbf{I}},\mathbf{d}\right) P_{c,\mathrm{FSK}}\left(\left\{ H_{j_n} \right\}_{\mathbf{J}},\left\{ f_{i_l} \right\}_{\mathbf{I}},\mathbf{d}\right), & \nonumber
\end{eqnarray}
\else
\begin{eqnarray}
 \label{Pc_cond}
 & P_c\left(\left\{ H_{j_n} \right\}_{\mathbf{J}},\left\{ f_{i_l} \right\}_{\mathbf{I}},\mathbf{d}\right) = P_{c,\mathrm{MPPM}}\left(\left\{ H_{j_n} \right\}_{\mathbf{J}},\left\{ f_{i_l} \right\}_{\mathbf{I}},\mathbf{d}\right)& \\
 & \cdot P_{c,\mathrm{OSSK}}\left(\left\{ H_{j_n} \right\}_{\mathbf{J}},\left\{ f_{i_l} \right\}_{\mathbf{I}},\mathbf{d}\right) P_{c,\mathrm{FSK}}\left(\left\{ H_{j_n} \right\}_{\mathbf{J}},\left\{ f_{i_l} \right\}_{\mathbf{I}},\mathbf{d}\right), & \nonumber
\end{eqnarray}
\fi
where, for the given FH-SMPPM symbol, the first term on the RHS is the probability of correctly detecting the MPPM symbol part, and the second and third terms on the RHS are the conditional probabilities of correctly detecting the OSSK and the FSK symbol parts, respectively, when the MPPM signal slots have been correctly identified.

It is clear that the OSSK and FSK probabilites do not depend on the specific MPPM pattern (specific positions of the signal slots within the FH-SMPPM symbol). Moreover, the OSSK probability does not depend on the specific FSK symbols, given that its detection is made using metric \eqref{metric_MPPM}, where the FSK part is cancelled out. Since the OSSK symbol sequence is produced independently at the transmitter, we can factorize the probability of correctly detecting the OSSK symbols, once the signal slots have been correctly identified, as
\begin{equation}
 \label{Pc_OSSK}
 P_{c,\mathrm{OSSK}}\left(\left\{ H_{j_n} \right\}_{\mathbf{J}}\right)=\prod_{n=0}^{w-1} \left(1-P_{e,\mathrm{OSSK}}\left( H_{j_n}\right)\right),
\end{equation}
where $P_{e,\mathrm{OSSK}}\left(H_{j_n}\right)$ is the symbol error probability associated to the detection of the $j_n$-th transmitter. As \eqref{metric_MPPM} is a Gaussian RV, it can be calculated in closed form as an instance of equally-spaced PAM signaling as \cite{Pro95}
\begin{eqnarray}
 \label{Pe_OSSK}
 & P_{e,\mathrm{OSSK}}\left(H_{j_n}\right) = \frac{1}{2} \mathrm{erfc}\left( \sqrt{\frac{T_s I_m^2 L_m^2}{4 N_0 \left(M_S-1\right)^2}} \right), \, j_n=0,M_S-1, & \nonumber\\
 & P_{e,\mathrm{OSSK}}\left(H_{j_n}\right) = \mathrm{erfc}\left( \sqrt{\frac{T_s I_m^2 L_m^2}{4 N_0 \left(M_S-1\right)^2}} \right), \, \mathrm{otherwise}. &
\end{eqnarray}
Notice that this probability is independent of the specific ordering of the OSSK symbol sequence, and its contribution to the overall symbol error probability can be calculated by taking into account the combinations with repetition of $w$ indexes taking values in $0,\cdots,M_S-1$. We can denote this set of combinations as $\mathcal{C}_{w}^{M_S}$, whose cardinality is $\binom{M_S+w-1}{w}$.

Respecting the probability of the FSK symbol part, given the symmetry of its constellation and the independence in the FSK symbol sequence, it will only depend on the received energy and we can write
\begin{equation}
 \label{Pc_FSK}
  P_{c,\mathrm{FSK}}\left(\left\{ H_{j_n} \right\}_{\mathbf{J}}\right)=\prod_{n=0}^{w-1} \left(1-P_{e,\mathrm{FSK}}\left( H_{j_n}\right)\right),
\end{equation}
where $P_{e,\mathrm{FSK}}\left( H_{j_n}\right)$ is the FSK symbol error probability when the $j_n$-th transmitter was active. This probability can be calculated as the average symbol error probability for non-coherent FSK when the received signal-to-noise ratio is $E_{s,\mathrm{FSK}}\left(H_{j_n}\right)/N_0$, namely \cite{Pro95}
\begin{equation}
 \label{Pe_FSK}
 P_{e,\mathrm{FSK}}\left(H_{j_n}\right)=\sum_{h=1}^{M_F-1} \frac{\left(-1\right)^{h-1}}{h+1} \binom{M_F-1}{h} \mathrm{e}^{- \frac{h}{\left( h+1 \right)} \frac{E_{s,\mathrm{FSK}}\left(H_{j_n}\right)}{N_0}},
\end{equation}
where $E_{s,\mathrm{FSK}}\left(H_{j_n}\right)$ is as given in \eqref{E_FSK}.

The derivation of the probability of correctly detecting the MPPM symbol is a bit more involved. Resorting to the ideas of \cite{Hamkins2005MultipulsePM} for the case of MPPM in the DCMC, we can calculate it as
\begin{equation}
 \label{MPPMpart}
 P_{c,\mathrm{MPPM}}\left(\left\{ H_{j_n} \right\}_{\mathbf{J}},\left\{ f_{i_l} \right\}_{\mathbf{I}},\mathbf{d}\right) = \int_{-\infty}^{\infty} p_{sl}\left(x\right) P_{nsl}\left(x\right) dx,
\end{equation}
where $x$ represents the minimum value attained by $X_k$ for the signal slots, $p_{sl}\left(x\right)$ is its probability density function (pdf), and $P_{nsl}\left(x\right)$ is the cumulative density function (cdf) of the $N-w$ non-signal slots, representing the probability that their $X_k$ values are lower or equal than $x$. As the RVs $X_k$ are zero-mean, Gaussian and independent from each other, it is straightforward to see that
\begin{equation}
 P_{nsl}\left(x\right) = \left(1-\frac{1}{2} \mathrm{erfc}\left(\frac{x}{\sqrt{N_0}} \right) \right)^{N-w}.
\end{equation}
The pdf $p_{sl}\left(x\right)$ can be calculated from its cdf $P_{sl}\left(x\right)$ \cite{Hamkins2005MultipulsePM}
\begin{equation}
 P_{sl}\left(x\right)=1-\prod_{n=0}^{w-1} \frac{1}{2} \mathrm{erfc}\left(\frac{x-\sqrt{T_s} H_{j_n}}{\sqrt{N_0}} \right),
\end{equation}
given that the RVs $X_k$ in this case are non-zero-mean, Gaussian and independent. This probability is calculated as one minus the probability that all the signal slots have $X_k>x$. Therefore,
\begin{equation}
 \displaystyle p_{sl}\left(x\right) = \sum_{n=0}^{w-1} \frac{1}{\sqrt{\pi N_0}} \mathrm{e}^{-\frac{\left( x-\sqrt{T_s} H_{j_n} \right)^2}{N_0}} \prod_{l=0,l\neq n}^{w-1} \frac{1}{2} \mathrm{erfc}\left(\frac{x-\sqrt{T_s} H_{j_l}}{\sqrt{N_0}} \right), 
\end{equation}
Notice that the resulting probability does not depend on the specific MPPM pattern $\mathbf{d}$, and, as $X_k$ does not depend on the FSK part, it can accordingly be denoted as $P_{c,\mathrm{MPPM}}\left(\left\{ H_{j_n} \right\}_{\mathbf{J}}\right)$. The expression for the probability of MPPM correct detection is, subsequently,
\ifonecol
\begin{equation}
\label{PcMPPM}
 P_{c,\mathrm{MPPM}}\left(\left\{ H_{j_n} \right\}_{\mathbf{J}}\right)=\displaystyle\sum_{n=0}^{w-1} \displaystyle\int_{-\infty}^{\infty} \frac{1}{\sqrt{\pi N_0}} \mathrm{e}^{-\frac{\left( x-\sqrt{T_s} H_{j_n} \right)^2}{N_0}} \prod_{l=0,l\neq n}^{w-1}  \frac{1}{2} \mathrm{erfc}\left(\frac{x-\sqrt{T_s} H_{j_l}}{\sqrt{N_0}} \right) \left(1-\frac{1}{2} \mathrm{erfc}\left(\frac{x}{\sqrt{N_0}} \right) \right)^{N-w} dx,
\end{equation}
\else
\begin{eqnarray}
\label{PcMPPM}
 & \displaystyle P_{c,\mathrm{MPPM}}\left(\left\{ H_{j_n} \right\}_{\mathbf{J}}\right)=\displaystyle\sum_{n=0}^{w-1} \displaystyle\int_{-\infty}^{\infty} \frac{1}{\sqrt{\pi N_0}} \mathrm{e}^{-\frac{\left( x-\sqrt{T_s} H_{j_n} \right)^2}{N_0}} \\
 &\displaystyle \cdot \prod_{l=0,l\neq n}^{w-1}  \frac{1}{2} \mathrm{erfc}\left(\frac{x-\sqrt{T_s} H_{j_l}}{\sqrt{N_0}} \right) \left(1-\frac{1}{2} \mathrm{erfc}\left(\frac{x}{\sqrt{N_0}} \right) \right)^{N-w} dx, \nonumber &
\end{eqnarray}
\fi
which has to be calculated numerically.

Taking into account that all these probabilities exclussively depend on the set of active transmitters and the corresponding received energies, but not on their specific ordering, the position of the signal slots within the MPPM symbol, or the FSK frequencies, we can write the final expression for the average symbol error probability of FH-SMPPM as
\ifonecol
\begin{equation}
 \label{Pe_final}
 \displaystyle P_e = 1 - \frac{1}{\binom{M_S+w-1}{w}} \sum_{\mathbf{J} \in \mathcal{C}_{w}^{M_S}} P_{c,\mathrm{MPPM}}\left(\left\{ H_{j_n} \right\}_{\mathbf{J}}\right) P_{c,\mathrm{OSSK}}\left(\left\{ H_{j_n} \right\}_{\mathbf{J}} \right) P_{c,\mathrm{FSK}}\left(\left\{ H_{j_n} \right\}_{\mathbf{J}} \right),
\end{equation}
\else
\begin{eqnarray}
 \label{Pe_final}
 & \displaystyle P_e = 1 - \frac{1}{\binom{M_S+w-1}{w}} \sum_{\mathbf{J} \in \mathcal{C}_{w}^{M_S}} P_{c,\mathrm{MPPM}}\left(\left\{ H_{j_n} \right\}_{\mathbf{J}}\right) & \nonumber \\
 & \displaystyle \cdot P_{c,\mathrm{OSSK}}\left(\left\{ H_{j_n} \right\}_{\mathbf{J}} \right) P_{c,\mathrm{FSK}}\left(\left\{ H_{j_n} \right\}_{\mathbf{J}} \right), &
\end{eqnarray}
\fi
where the different probabilities of correct detection are as given in \eqref{PcMPPM}, \eqref{Pc_OSSK} and \eqref{Pc_FSK}, respectively.

\subsubsection{Average bit error probability}
\label{ABEP}

Following the same ideas in order to calculate the average bit error probability, we can average over the conditional bit error probability, so that
\ifonecol
\begin{equation}
 \label{Pb_avg}
 \displaystyle P_b = \mathrm{E}\left[P_b\left(\left\{ H_{j_n} \right\}_{\mathbf{J}},\left\{ f_{i_l} \right\}_{\mathbf{I}},\mathbf{d}\right)\right]= \frac{1}{\binom{M_S+w-1}{w}} \sum_{\mathbf{J} \in \mathcal{C}_{w}^{M_S}}  P_b\left(\left\{ H_{j_n} \right\}_{\mathbf{J}}\right),
\end{equation}
\else
\begin{eqnarray}
 \label{Pb_avg}
 & \displaystyle P_b = \mathrm{E}\left[P_b\left(\left\{ H_{j_n} \right\}_{\mathbf{J}},\left\{ f_{i_l} \right\}_{\mathbf{I}},\mathbf{d}\right)\right]& \\
 & \displaystyle = \frac{1}{\binom{M_S+w-1}{w}} \sum_{\mathbf{J} \in \mathcal{C}_{w}^{M_S}}  P_b\left(\left\{ H_{j_n} \right\}_{\mathbf{J}}\right),& \nonumber
\end{eqnarray}
\fi
where we have already taken into account that, given the analysis made while developing the calculation of $P_e$, the conditional probability will only depend on the specific OSSK pattern. It can be factorized under the mutually exclusive hypothesis of correct and erroneous demodulation of MPPM, so that
\ifonecol
\begin{eqnarray}
 &P_b\left(\left\{ H_{j_n} \right\}_{\mathbf{J}}\right) = p_b\left(\left\{ H_{j_n} \right\}_{\mathbf{J}} \big\vert c, \mathrm{MPPM} \right) P_{c,\mathrm{MPPM}}\left(\left\{ H_{j_n} \right\}_{\mathbf{J}}\right) & \\
 & + p_b\left(\left\{ H_{j_n} \right\}_{\mathbf{J}} \big\vert e, \mathrm{MPPM} \right) \left( 1 -  P_{c,\mathrm{MPPM}}\left(\left\{ H_{j_n} \right\}_{\mathbf{J}}\right)\right), & \nonumber
\end{eqnarray}
\else
\begin{eqnarray}
 & P_b\left(\left\{ H_{j_n} \right\}_{\mathbf{J}}\right) & \\
 & = p_b\left(\left\{ H_{j_n} \right\}_{\mathbf{J}} \big\vert c, \mathrm{MPPM} \right) P_{c,\mathrm{MPPM}}\left(\left\{ H_{j_n} \right\}_{\mathbf{J}}\right) & \nonumber \\
 & + p_b\left(\left\{ H_{j_n} \right\}_{\mathbf{J}} \big\vert e, \mathrm{MPPM} \right) \left( 1 -  P_{c,\mathrm{MPPM}}\left(\left\{ H_{j_n} \right\}_{\mathbf{J}}\right)\right), & \nonumber
\end{eqnarray}
\fi
where $p_b\left(\cdot,\cdot \vert \cdot \right)$ is the proportion of erroneous bits under the given hypothesis (i.e. correct or incorrect detection of the MPPM symbol). The value $p_b\left(\left\{ H_{j_n} \right\}_{\mathbf{J}} \big\vert c, \mathrm{MPPM} \right)$ is the proportion of bits in error in the demodulation of the OSSK and FSK symbol parts when the MPPM demodulation has correctly identified the signal slots, and only the errors in demodulating OSSK and FSK have to be taken into account. We can calculate it as
\begin{equation}
 p_b\left(\left\{ H_{j_n} \right\}_{\mathbf{J}} \big\vert c, \mathrm{MPPM} \right) = \frac{ne_{\mathrm{OSSK}}+ne_{\mathrm{FSK}}}{q_{\mathrm{F-S}}},
\end{equation}
where $ne_{\mathrm{OSSK}}$ is the average number of erroneous bits determined by the detection of the OSSK symbols, and $ne_{\mathrm{FSK}}$ is the average number of erroneous bits determined by the detection of the FSK symbols. They can be added as such because the detection process of both kind of symbols is fully independent once the signal slots have been decided (i.e. after the detection of MPPM).

In the case of OSSK, as the mapping from bits to symbols is gray, we can approximate the bit error probability associated to the symbol corresponding to $H_{j_n}$ as $P_{e,\mathrm{OSSK}}\left(H_{j_n}\right)/n_S$. This is the proportion of erroneous bits given by an error in detecting this OSSK symbol, and $n_S \cdot P_{e,\mathrm{OSSK}}\left(H_{j_n}\right)/n_S$ will be its contribution to the total within the FH-SMPPM symbol. As we have a set of $w$ OSSK symbols,
\begin{equation}
 ne_{\mathrm{OSSK}} = \sum_{n=0}^{w-1} P_{e,\mathrm{OSSK}}\left(H_{j_n}\right),
\end{equation}
where the conditional symbol error probability can be approximated using \eqref{Pe_OSSK}. In the case of FSK, a similar derivation leads to
\begin{equation}
 ne_{\mathrm{FSK}} = n_F \frac{2^{n_F-1}}{2^{n_F}-1} \sum_{n=0}^{w-1} P_{e,\mathrm{FSK}}\left(H_{j_n}\right),
\end{equation}
where we have taken into account that the proportion of erroneous bits for an FSK symbol error is estimated as $2^{n_F-1}/\left(2^{n_F}-1\right)$ \cite{Pro95}, and the conditional symbol error probability can be approximated by \eqref{Pe_FSK}. Notice that the average number of erroneous bits given are quantities that depend on the specific OSSK pattern, and are conditional measurements to be later jointly averaged following \eqref{Pb_avg}.

For the complementary hypothesis, we have
\ifonecol
\begin{equation}
 \displaystyle p_b\left(\left\{ H_{j_n} \right\}_{\mathbf{J}} \big\vert e, \mathrm{MPPM} \right) = \frac{ne_{\mathrm{MPPM}}+ne_{\mathrm{OSSK}}^{sl}+ne_{\mathrm{OSSK}}^{nsl}+ne_{\mathrm{FSK}}^{sl}+ne_{\mathrm{FSK}}^{nsl}}{q_{\mathrm{F-S}}},
\end{equation}
\else
\begin{eqnarray}
 &\displaystyle p_b\left(\left\{ H_{j_n} \right\}_{\mathbf{J}} \big\vert e, \mathrm{MPPM} \right)& \\
 &\displaystyle = \frac{ne_{\mathrm{MPPM}}+ne_{\mathrm{OSSK}}^{sl}+ne_{\mathrm{OSSK}}^{nsl}+ne_{\mathrm{FSK}}^{sl}+ne_{\mathrm{FSK}}^{nsl}}{q_{\mathrm{F-S}}},& \nonumber
\end{eqnarray}
\fi
where $ne_{\mathrm{MPPM}}$ is the average number of erroneous bits in the demodulation of MPPM, $ne_{\mathrm{OSSK}}^{sl}$ is the average number of erroneous bits in the demodulation of OSSK for the proportion of correctly identified signal slots, $ne_{\mathrm{OSSK}}^{nsl}$ is the average number of erroneous bits when applying OSSK demodulation to the non-signal slots erroneously identified as signal slots, and $ne_{\mathrm{FSK}}^{sl}$ and $ne_{\mathrm{FSK}}^{nsl}$ are the corresponding average number of errors in the FSK symbol part. The estimated proportion of bits affected by an MPPM detection error is $2^{q_{\mathrm{MPPM}}-1}/\left(2^{q_{\mathrm{MPPM}}}-1\right)$ \cite{a029b3b4b02e4b40ac8ca72cc4f7fc8c}, so that the corresponding average number of erroneous bits will be
\begin{equation}
\label{eq_ne_MPPM}
 ne_{\mathrm{MPPM}} = q_{\mathrm{MPPM}} \frac{2^{q_{\mathrm{MPPM}}-1}}{2^{q_{\mathrm{MPPM}}}-1}.
\end{equation}

For $ne_{sl,\mathrm{OSSK}}$ we have an average number of erroneous bits per detected OSSK symbol in a signal slot of $P_{e,\mathrm{OSSK}}\left(H_{j_n}\right)$, and now we have to take into account the average number of signal slots correctly identified. This can be calculated as \cite{6620996}
\begin{equation}
 \frac{\sum\limits_{l=1}^{\min\left(w,N-w\right)} {\binom{w}{l}} {\binom{N-w}{l}} \left(w-l\right) }{\binom{N}{w}-1} = \sum_{l=1}^{\min\left(w,N-w\right)} K_l \left(w-l\right),
\end{equation}
where the index $l$ is the number of signal slots missed in the detection of MPPM, and we have defined $K_l={\binom{w}{l}} {\binom{N-w}{l}}/\left({\binom{N}{w}-1}\right)$. Consequently,
\begin{equation}
 ne_{\mathrm{OSSK}}^{sl} = \sum\limits_{l=1}^{\min\left(w,N-w\right)} K_l \left(w-l\right) \frac{1}{w} \sum_{n=0}^{w-1} P_{e,\mathrm{OSSK}}\left(H_{j_n}\right).
\end{equation}
In the case of $ne_{\mathrm{OSSK}}^{nsl}$, we can make the reasonable assumption that on average half of the bits involved in the demodulation of OSSK over a non-signal slot will be in error, so that
\begin{equation}
 ne_{\mathrm{OSSK}}^{nsl} =  \frac{n_S}{2} \sum\limits_{l=1}^{\min\left(w,N-w\right)} K_l \, l,
\end{equation}
In the same way, we can develop the expressions for the average number of errors of FSK as
\ifonecol
\begin{equation}
  ne_{\mathrm{FSK}}^{sl} = n_F \frac{2^{n_F-1}}{2^{n_F}-1} \sum\limits_{l=1}^{\min\left(w,N-w\right)} K_l \left(w-l\right) \frac{1}{w} \sum_{n=0}^{w-1} P_{e,\mathrm{FSK}}\left(H_{j_n}\right),
\end{equation}
\else
\begin{eqnarray}
 &\displaystyle ne_{\mathrm{FSK}}^{sl} & \\
 &\displaystyle =n_F \frac{2^{n_F-1}}{2^{n_F}-1} \sum\limits_{l=1}^{\min\left(w,N-w\right)} K_l \left(w-l\right) \frac{1}{w} \sum_{n=0}^{w-1} P_{e,\mathrm{FSK}}\left(H_{j_n}\right),& \nonumber
\end{eqnarray}
\fi
and
\begin{equation}
 ne_{\mathrm{FSK}}^{nsl} =  \frac{n_F}{2} \sum\limits_{l=1}^{\min\left(w,N-w\right)} K_l \, l.
\end{equation}
Notice that we are implicitly assuming that all the possible MPPM patterns in $\mathcal{S}_{\mathrm{MPPM}}$ excepting the hypothetical $\mathbf{d}\in \mathcal{S}_{\mathrm{MPPM}}^*$ can be chosen in the demodulation. Therefore, the previous expressions will only be exact if $\log_2\binom{N}{w}$ is integer. As normally this is not the case, the results should be then interpreted as approximations, but, given that the difference between $\log_2\binom{N}{w}$ and $\floor*{\log_2\binom{N}{w}}$ is in practice small, the mismatch will not be significant, specially for high signal-to-noise ratios.

Taking all this into account, the final expression for the average bit error probability can be written as
\ifonecol
 \begin{eqnarray}
 \label{Pb_avg_tot}
 &\displaystyle P_b = \frac{1}{q_{\mathrm{F-S}}} \frac{1}{\binom{M_S+w-1}{w}} \sum_{\mathbf{I}\in \mathcal{C}_{w}^{M_S}} \!\left\{ P_{c,\mathrm{MPPM}}\left(\left\{ H_{j_n} \right\}_{\mathbf{J}}\right) \sum_{n=0}^{w-1} \left( P_{e,\mathrm{OSSK}}\left(H_{j_n}\right) + n_F \frac{2^{n_F-1}}{2^{n_F}-1} P_{e,\mathrm{FSK}}\left(H_{j_n}\right) \right) \right. \nonumber \\
 &\displaystyle \left. + \left( 1- P_{c,\mathrm{MPPM}}\left(\left\{ H_{j_n} \right\}_{\mathbf{J}}\right) \right) \cdot \Bigg[q_{\mathrm{MPPM}} \frac{2^{q_{\mathrm{MPPM}}-1}}{2^{q_{\mathrm{MPPM}}}-1}\right. & \nonumber\\
 &\displaystyle \left. + \sum_{l=1}^{\min\left(w,N-w\right)} K_l \left( \left(w-l\right) \frac{1}{w} \sum_{n=0}^{w-1} \left( P_{e,\mathrm{OSSK}}\left(H_{j_n}\right) + n_F \frac{2^{n_F-1}}{2^{n_F}-1} P_{e,\mathrm{FSK}}\left(H_{j_n}\right) \right) + \frac{n_S+n_F}{2} l \right) \Bigg] \right\}, &
\end{eqnarray}
\else
 \begin{eqnarray}
 \label{Pb_avg_tot}
 &\displaystyle P_b = \frac{1}{q_{\mathrm{F-S}}} \frac{1}{\binom{M_S+w-1}{w}} \sum_{\mathbf{I}\in \mathcal{C}_{w}^{M_S}} \Biggl\{ P_{c,\mathrm{MPPM}}\left(\left\{ H_{j_n} \right\}_{\mathbf{J}}\right) & \nonumber \\
 &\displaystyle \cdot \sum_{n=0}^{w-1} \left( P_{e,\mathrm{OSSK}}\left(H_{j_n}\right) + n_F \frac{2^{n_F-1}}{2^{n_F}-1} P_{e,\mathrm{FSK}}\left(H_{j_n}\right) \right) & \nonumber \\
 &\displaystyle + \left( 1- P_{c,\mathrm{MPPM}}\left(\left\{ H_{j_n} \right\}_{\mathbf{J}}\right) \right)  \Biggl[ q_{\mathrm{MPPM}} \frac{2^{q_{\mathrm{MPPM}}-1}}{2^{q_{\mathrm{MPPM}}}-1} & \nonumber\\
 &\displaystyle + \sum_{l=1}^{\min\left(w,N-w\right)} K_l \Biggl( \left(w-l\right) \frac{1}{w} \sum_{n=0}^{w-1} \biggl( P_{e,\mathrm{OSSK}}\left(H_{j_n}\right) & \nonumber \\
 &\displaystyle + n_F \frac{2^{n_F-1}}{2^{n_F}-1} P_{e,\mathrm{FSK}}\left(H_{j_n}\right) \biggr) + \frac{n_S+n_F}{2} l \Biggr) \Biggr] \Biggr\}, &
\end{eqnarray}
\fi
and has to be calculated numerically.

\section{Simulation results}
\label{results}

In this Section we present simulation results for the BER and SER, along the calculated values for $P_b$ and $P_e$ using the previously developed formulas\footnote{The numerical calculation of the integrals in the formulas in the previous Section has been performed using standard functions for numerical integration as available in mathematical software packages like Matlab(R).}. For comparison, we offer results for SMPPM \cite{8647138} and I-TFH \cite{8720057}, as limit cases where either the FSK symbol part or the OSSK symbol part have been removed. Notice that the authors of \cite{8647138} only analyzed the average symbol error probability $P_e$ of SMPPM. We will give results for both the average symbol and average bit error probability of SMPPM using equations \eqref{Pe_final} and \eqref{Pb_avg_tot}, respectively, by setting $P_{e,\mathrm{FSK}}\left(\cdot\right)=0$. To estimate the SER/BER, the detection of I-TFH is done as proposed in \cite{8720057}, while the detection of SMPPM is done as suggested by the hypothesis leading to \eqref{Pe_final} and \eqref{Pb_avg_tot}: first the MPPM signal slots are identified, and then the OSSK detector is applied to the signals in such slots.

\begin{figure}[htb!]
\centering
\includegraphics[width=\figurewidth, keepaspectratio]{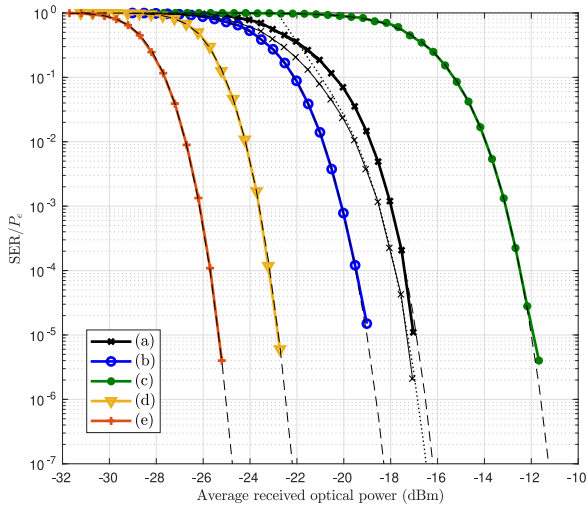}
 \caption{Average SER and calculated $P_e$ for I-TFH, SMPPM and FH-SMPPM with the same number of bits per symbol. In all the cases, $N=8$, $w=4$ and $R_b=100$ Mbps. (a) FH-SMPPM (bold black), $M_F=M_S=4$, $L_m=0.7$, $m=0.5$. The simulation results for coherent detection of FSK with the same parameters are plotted with a thin black curve, and the corresponding union bound with a thin dotted black curve. (b) FH-SMPPM, same parameters excepting $m=0.9$. (c) SMPPM. $M_S=16$ and $L_m=0.7$. (d) I-TFH, $M_F=16$, $m=0.5$. (e) I-TFH, same parameters excepting $m=0.9$. SER is represented in solid color lines, calculated $P_e$ is represented in dashed black lines.} \label{fig6}
\end{figure}
For better comparison with the state-of-the-art, as we are in the OWC domain, we present the results with respect to the average received optical power $P_{opt}$ (see \eqref{arop}), for a given bit rate $R_b$. To this end, we will take the typical parameter values shown in Table \ref{Tab2}. To obtain the average received optical power, first we compute the value of $I_{DC}$ for a given noise density $N_0$ solving the quadratic equation given by \eqref{N0}. Such value is then plugged into equation \eqref{arop} to get the corresponding average optical power value.
\begin{table}
\begin{center}
\begin{tabular}{|l|l|}
  \hline
   Reference absolute temperature   & $T_R=290$ K \\
   \hline
   PD load resistor & $R_L=50$ $\Omega$\\
   \hline
   Noise figure of the receiver electronics & $NF=10\log_{10}\left(F\right)=10$ dB\\
   \hline
   Relative-intensity noise factor   & $(RIN)=-155$ dB/Hz\\
   \hline
   PD responsivity &$\mathcal{R}=0.5$ A/W.\\
   \hline
\end{tabular}
 \vspace*{0.2cm}
 \caption{Physical parameter values for the simulations.}
 \label{Tab2}
\end{center}
\end{table}

\begin{figure}[htb!]
\centering
\includegraphics[width=\figurewidth, keepaspectratio]{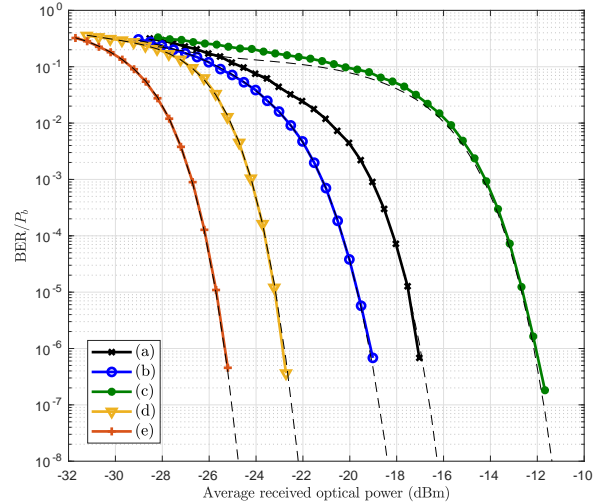}
 \caption{Average BER and calculated $P_b$ for I-TFH, SMPPM and FH-SMPPM with the same number of bits per symbol. In all the cases, $N=8$, $w=4$ and $R_b=100$ Mbps. (a) FH-SMPPM, $M_F=M_S=4$, $L_m=0.7$, $m=0.5$. (b) FH-SMPPM, same parameters excepting $m=0.9$. (c) SMPPM. $M_S=16$ and $L_m=0.7$. (d) I-TFH, $M_F=16$, $m=0.5$. (e) I-TFH, same parameters excepting $m=0.9$. BER is represented in solid color lines, calculated $P_b$ is represented in dashed black lines.} \label{fig7}
\end{figure}
In Figs. \ref{fig6} and \ref{fig7}, we can see the simulation results and calculated average error probabilities for some cases of interest of SMPPM, I-TFH and FH-SMPPM. In Fig. \ref{fig6}, we have depicted the symbol error rate (SER) and the calculated average symbol error probability ($P_e$). The $P_e$ for I-TFH  is calculated based on the derivations from \cite{8720057}, and the $P_e$ for SMPPM and FH-SMPPM is calculated from \eqref{Pe_final}. In Fig. \ref{fig7}, we have the corresponding results for the BER and the average bit error probability ($P_b$): again, $P_b$ for I-TFH is based on \cite{8720057}, and $P_b$ for SMPPM and FH-SMPPM is calculated from \eqref{Pb_avg_tot}. We have also provided in Fig. \ref{fig6} the union bound (UB) for the ML detection of FH-SMPPM, along the simulation results with coherent FSK, for the parameters of case a). These curves verify that our proposed detector is close to optimum in the ML sense excepting for the non-coherence of FSK. Notice that the complexity of implementing a coherent FSK detector usually offsets the gain in SNR, which may be less than $1$ dB.

As we may see, in all the cases, the formulas for $P_e$ and $P_b$ provide an excellent agreement with respect to the actual SER and BER simulated values, thus validating the developments of the previous Section. Only the $P_b$ for SMPPM slightly diverges from the actual BER in the lowest received power region, due to the approximations made in the calculations, but the difference is far from significant. As for the differences among the behaviour of the different systems, we can verify that SMPPM performs poorer than both I-TFH and FH-SMPPM for the same total number of bits mapped into the OSSK and/or FSK parts (SMPPM maps $4w$ bits into the OSSK symbol part, FH-SMPPM maps $2w$ bits into the OSSK symbol part and $2w$ bits into the FSK symbol part, and I-TFH maps $4w$ bits into the FSK symbol part).

FH-SMPPM is thus a system that provides error rate results that lie in between I-TFH and SMPPM: the comparative advantage with respect to SMPPM lies in the fact that the information plugged into the FSK symbols is better protected, but at the cost of higher bandwidth expenditure. This is a known result for FSK-based modulations against amplitude-based modulations like OSSK, where an increase in efficiency leads to poorer distance properties and hence to a degradation in error rate performance. Conversely, the comparative advantage with respect to I-TFH consists in a relatively higher spectral efficiency, at the cost of worse error rate results. In Figs. \ref{fig6} and \ref{fig7}, we can also see that, for FH-SMPPM, the modulation index $m$ plays an important role, since it is directly related to the final error rate of the FSK part. This is also the case for I-TFH, as shown in the same figures, and as it was already ascertained in \cite{8720057}.

\begin{figure}[htb!]
\centering
\includegraphics[width=\figurewidth, keepaspectratio]{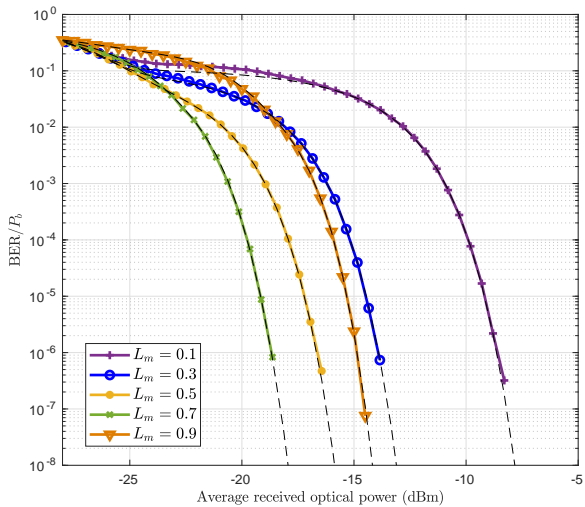}
 \caption{Average BER and calculated $P_b$ for FH-SMPPM with $N=16$, $w=8$, $M_F=8$, $M_S=4$, $m=0.9$, and different $L_m$ values. Binary data rate is $R_b=75$ Mbps. BER is represented in solid color lines, calculated $P_b$ is represented in dashed black lines.} \label{fig8}
\end{figure}
The role of $L_m$ (the span in the OSSK part) can be comparatively verified in Fig. \ref{fig8}, and it follows the trends already identified in previous works \cite{6241395,8647138}. It can be seen that it is convenient to set a value in between the extreme possibilities (either $0$ or $1$): the OSSK part behaves better for the largest span possible ($L_m$ far from $0$), but, if $L_m \rightarrow 1$, there will be signal slots with energy close to zero. These become almost impossible to identify as such at the receiver end, so that the error rate in the MPPM part degrades, and the corresponding information bits carried over the FSK part of the missed signal slot will be directly lost. This can be verified in the figure, where there is a steady improvement from $L_m=0.1$ to $L_m=0.7$, but $L_m=0.9$ leads to poorer results. On the other hand, just as before, the calculated $P_b$ shows to be in excellent agreement with the simulated BER.

\begin{table*}
\begin{center}
  \begin{tabular}{|l|c|c|c|c|c|}
  \hline
   & $P_{opt}$ & Power & Spectral effi-  & Complexity & Latency \\
   & (dBm) & efficiency & ciency (bps/Hz) & (ops per bit) & ($\times T_s$) \\
  \hline
  (a) FH-SMPPM $M_F=M_S=4$, $L_m=0.7$, $m=0.5$ & $-17.0$ & $0.056$ & $0.55$ & $67.4$ & $9$ \\
  \hline
  (b) FH-SMPPM $M_F=M_S=4$, $L_m=0.7$, $m=0.9$ & $-19.0$ & $0.109$ & $0.55$ & $67.4$ & $9$\\
  \hline
  (c) SMPPM $M_S=4$, $L_m=0.7$ & $-11.9$ & $0.006$ & $1.38$ & $4.9$ & $9$ \\
  \hline
  (d) I-TFH $M_F=16$, $m=0.5$ & $-22.8$ & $0.306$ & $0.16$  & $819.4$ & $9$ \\
  \hline
  (e) I-TFH $M_F=16$, $m=0.9$ & $-25.3$ & $0.793$ & $0.16$ & $819.4$ & $9$ \\
  \hline
  \end{tabular}
  \vspace*{0.2cm}
\caption{Different performance values for the systems represented in Figs. \ref{fig6} and \ref{fig7}. The alphanumeric key represents the analogous case in said figures. The average received optical power $P_{opt}$ is given for a target BER of $10^{-5}$.}
\label{Tab3}
\end{center}
\end{table*}
To further illustrate the properties of FH-SMPPM, we show in Table \ref{Tab3} different performance measurements for the systems whose error rate is represented in Figs. \ref{fig6} and \ref{fig7}. We have used the expressions of Section \ref{analysis}, by zeroing the corresponding contributions when considering the limit cases SMPPM and I-TFH. We can clearly identify the trends previously mentioned: I-TFH performs poorer in terms of complexity (it requires far more resources) and spectral efficiency, but the error rate performance and the power efficiency improves dramatically with respect to pure SMPPM. These comparative properties can be traded-off by using FH-SMPPM and setting $M_F$ and $M_S$ values according to pre-defined performance goals of different kinds. We can thus verify how FH-SMPPM constitutes a flexible system with several degrees of freedom that can efficiently trade-off the advantages and disadvantages inherent to both I-TFH and SMPPM. This happens not only in terms of the average error rates, but also from the point of view of spectral/power efficiencies and implementation complexity. Notice that the latency in Table \ref{Tab2} remains constant as $N$ is constant. Additionally, MPPM parameters $N$ and $w$ can be further tuned to reach target performances with given resources.

\section{Conclusions}
\label{conclusions}

In this article we have introduced a new index modulated waveform, denominated FH-SMPPM, where the information bits can be mapped along three dimensions: time, space and frequency. We have analyzed its performance in terms of spectral and power efficiencies, the general complexity of the demodulator, and we have developed formulas for the average symbol and bit error probabilities in the non-turbulent free-space optical channel. The calculated average error probabilities provide an excellent match to the simulation results, thus validating the analytical approach unfolded. To the best of our knowledge, this is the first work dealing in depth with this three-dimensional IM system, and it provides solid tools and guidelines for its practical application and/or for the derivation of other related alternatives.

FH-SMPPM possesses properties that makes it suitable for the optical wireless communication environment, since its baseline is pulsed-position modulation. The comparison with other MPPM-based related alternatives in the turbulence-free FSO channel, I-TFH and SMPPM, shows that FH-SMPPM provides results that lie in between the performance of both alternatives. I-TFH is the extreme case of FH-SMPPM when the spatial dimension is not present, and it is better performing in SER/BER, and SMPPM is the extreme case of FH-SMPPM when the frequency dimension is not driven, and it is worse performing in SER/BER. As an intermediate alternative, driving both frequency and space, FH-SMPPM can be configured to trade-off the advantages and disadvantages of I-TFH and SMPPM in a very convenient way. Future studies will ascertain its capabilities in time-varying optical channels, where the flexibility of the IM waveform could be easily turned into an asset.

\section*{Acknowledgments}

The authors would like to acknowledge finantial support from the Canada research chair Tier 2.


\ifCLASSOPTIONcaptionsoff
  \newpage
\fi



\bibliographystyle{IEEEtran}
%
%
%

%

\begin{IEEEbiography}[{\includegraphics[width=1in,height=1.5in,clip,keepaspectratio]{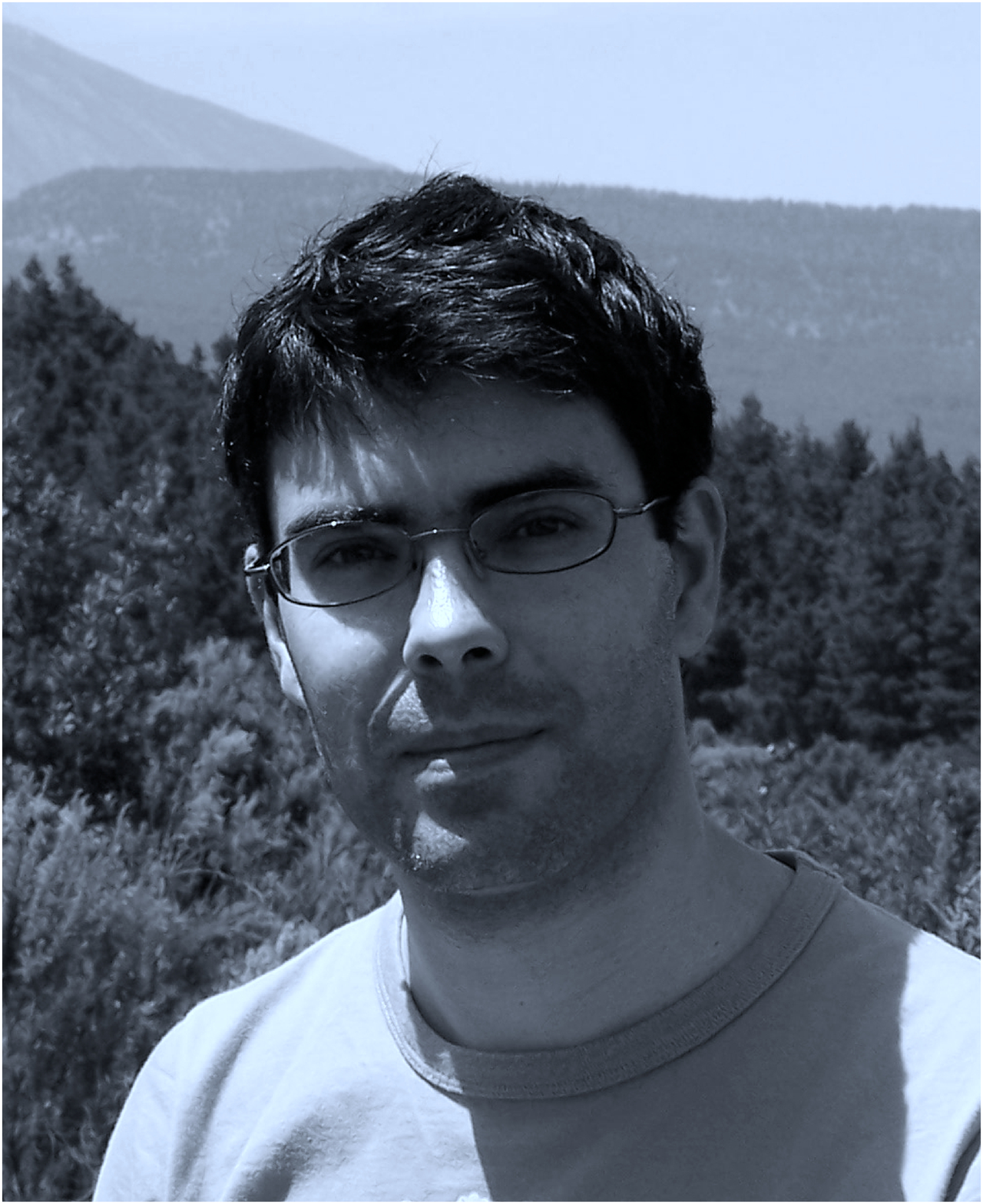}}]
{Francisco J. Escribano} (M'06, SM'16) received his degree in Telecommunications Engineering at ETSIT-UPM, Spain, and his Ph.D. degree at Universidad Rey Juan Carlos, Spain. He is currently Associate Professor at the Department of Signal Theory \& Communications of Universidad de Alcal\'{a}, Spain, where he is involved in several undergraduate and master courses in Telecommunications Engineering. He has been Visiting Researcher at the Politectnico di Torino, Italy, and at the EPFL, Switzerland. His research activities are focused on Communications Systems and Information Theory, mainly on the topics of channel coding, modulation and multiple access, and on the applications of Chaos in Engineering.
\end{IEEEbiography}

\begin{IEEEbiography}[{\includegraphics[width=1in,height=1.25in,clip,keepaspectratio]{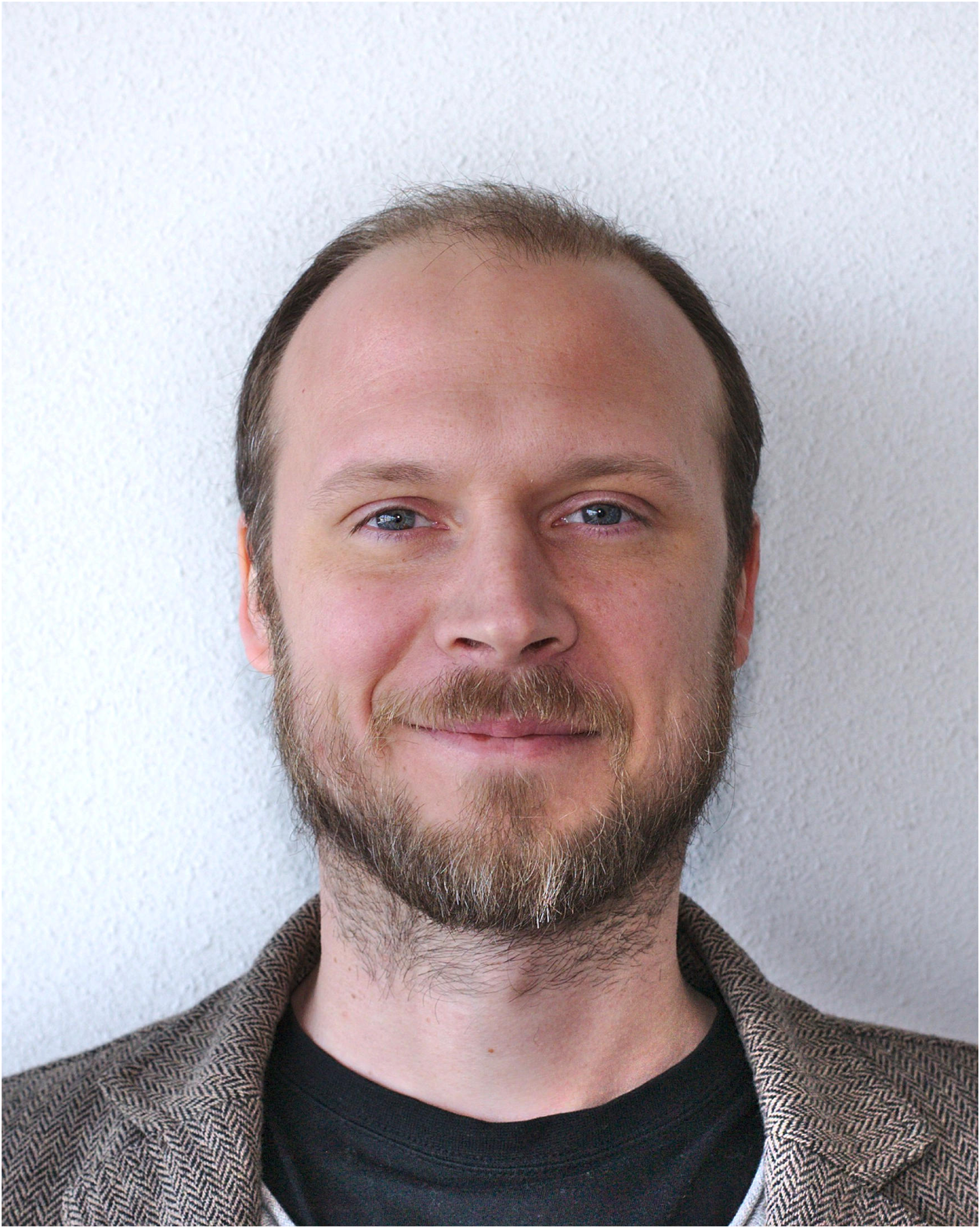}}]
{Alexandre Wagemakers} received the Telecommunication Engineering degree in 2003 from the Polytechnic University in Madrid, Spain, and the Ph.D. degree in Applied Physics in 2008 from the University Rey Juan Carlos, Madrid, Spain, where he is working as an Assistant Professor in the Department of Physics. His interests are currently the theory and applications of Nonlinear Dynamics.
\end{IEEEbiography}

\begin{IEEEbiography}[{\includegraphics[width=1in,height=1.25in,clip,keepaspectratio]{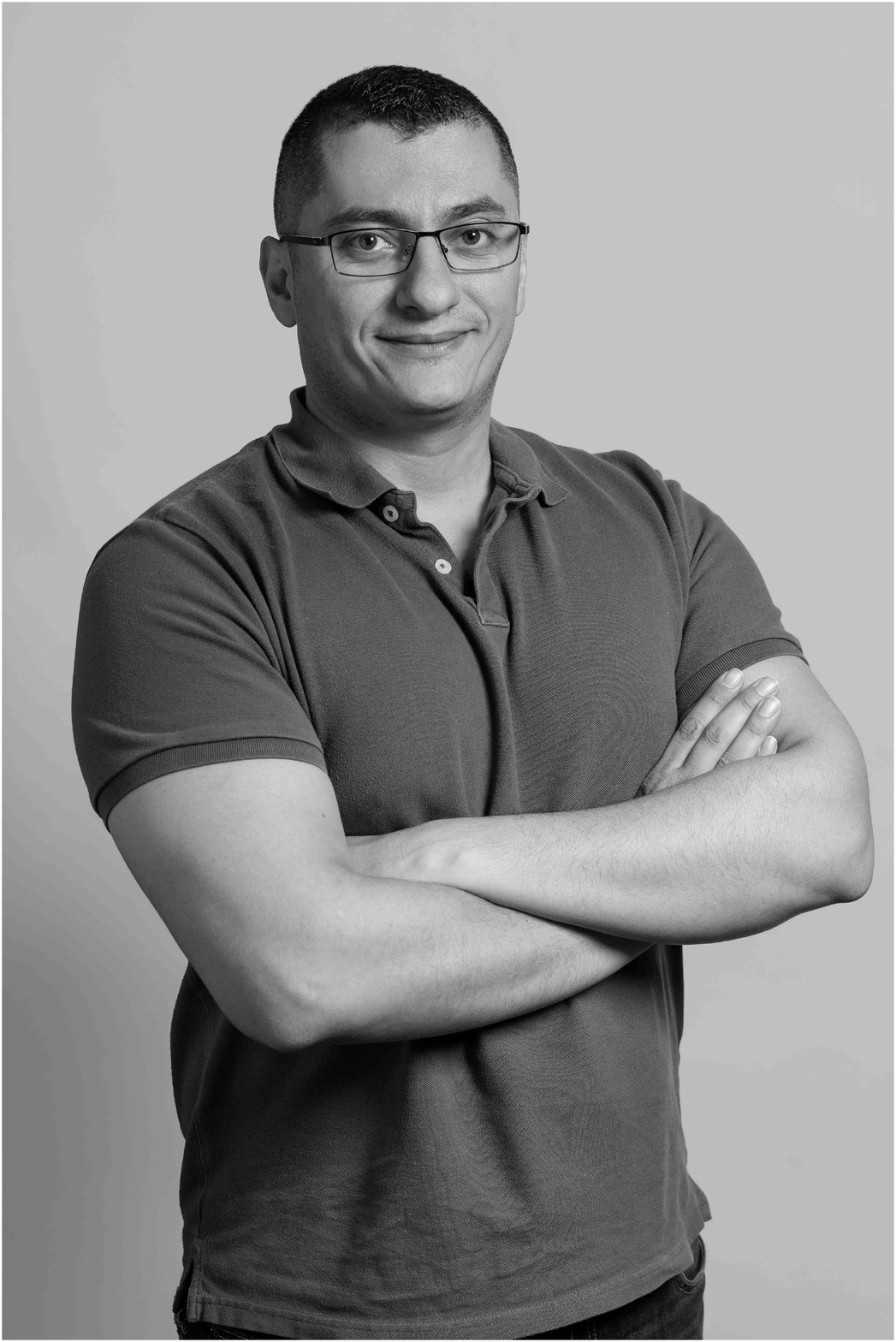}}]
{Georges Kaddoum} received the Bachelor’s degree in electrical engineering from the \'{E}cole Nationale Sup\'{e}rieure de Techniques Avanc\'{e}es (ENSTA Bretagne), Brest, France, and the M.S. degree in telecommunications and signal processing(circuits, systems, and signal processing) from the Universit\'{e} de Bretagne Occidentale and Telecom Bretagne (ENSTB), Brest, in 2005 and the Ph.D. degree (with honors) in signal processing and telecommunications from the National Institute of Applied Sciences (INSA), University of Toulouse, Toulouse, France, in 2009. He is currently an Associate Professor and Tier 2 Canada Research Chair with the \'{E}cole de Technologie Sup\'{e}rieure (\'{E}TS), Universit\'{e} du Qu\'{e}bec, Montr\'{e}al, Canada. He was awarded the \'{E}TS Research Chair in physical-layer security for wireless networks in 2014, and the prestigious Tier 2 Canada Research Chair in wireless IoT networks in 2019. Since 2010, he has been a Scientific Consultant in the field of space and wireless telecommunications for several US and Canadian companies. He has published over 150+ journal and conference papers and has two pending patents. His recent research activities cover mobile communication systems, modulations, security, and space communications and navigation. Dr. Kaddoum received the Best Papers Awards at the 2014 IEEE International Conference on Wireless and Mobile Computing, Networking, Communications (WIMOB), with three co-authors, and at the 2017 IEEE International Symposium on Personal Indoor and Mobile Radio Communications (PIMRC), with four co-authors. Moreover, he received IEEE Transactions on Communications Exemplary Reviewer Award for the year 2015 and 2017. In addition, he received the research excellence award of the Universit\'{e} du Qu\'{e}bec in the year 2018. In the year 2019, he received the research excellence award from the \'{E}TS in recognition of his outstanding research outcomes. Prof. Kaddoum is currently serving as an Associate Editor for IEEE Transactions on Information Forensics and Security, and IEEE Communications Letters.
\end{IEEEbiography}

\begin{IEEEbiography}[{\includegraphics[width=1in,height=1.25in,clip,keepaspectratio]{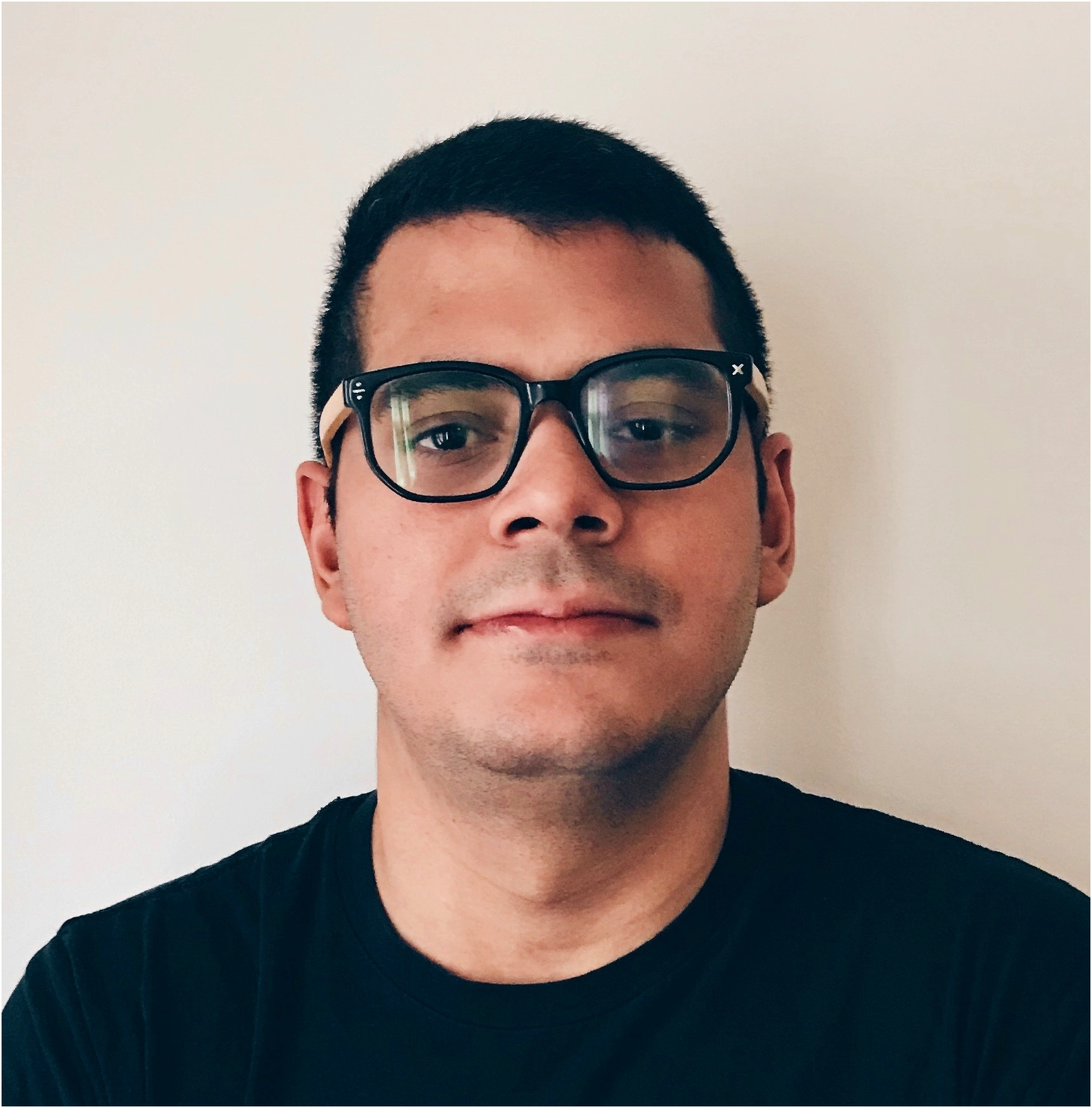}}]
{Joao V. C. Evangelista} received the B.S. and M.S. degree in electrical engineering from the Universidade Federal de Pernambuco, Recife, Brazil, in 2015 and 2016 respectively. In 2016, he joined the electrical engineering department at the \'{E}cole de Technologie Sup\'{e}rieure (\'{E}TS) as a Ph.D. student. He was awarded the Mitacs Globalink Fellowship in 2016 and the Fonds de recherche du Qu\'{e}bec - Nature et technologies Doctoral Fellowship in 2019. His current research interests include non-orthogonal multiple access communications, machine-to-machine communications and stochastic geometric modeling of wireless networks.
\end{IEEEbiography}








\end{document}